\documentclass[prb,amsmath,amssymb,amsbsy,twocolumn,superscriptaddress,showpacs]{revtex4-1}

\usepackage{amsfonts} 
\usepackage{amsmath}
\usepackage{amssymb}
\usepackage{cancel} 
\usepackage{bm}
\usepackage[usenames, dvipsnames]{color} 
\usepackage{dcolumn}
\usepackage{epic} 
\usepackage{epsfig}
\usepackage{eucal} 
\usepackage{graphicx}
\usepackage[breaklinks,colorlinks = true,linkcolor = red,urlcolor=cyan,citecolor=red]{hyperref}
\usepackage{mathrsfs}
\usepackage{soul}
\usepackage{subfigure}
\usepackage{varwidth}
\usepackage{wrapfig} 
\usepackage{xy} 
\usepackage{times}
\usepackage{bm}

\begin{document}
\newcommand{\bij}[3]{(b^{#1}_{#2}b^{#1}_{#3})}
\newcommand{\bii}[3]{(b^{#2}_{#1}b^{#3}_{#1})}
\newcommand{\uij}[2]{u^{#1}_{#2}}

\title{Heisenberg-Kitaev model on hyperhoneycomb lattice}

\author{Eric Kin-Ho Lee}
\author{Robert Schaffer}
\author{Subhro Bhattacharjee}
\affiliation{Department of Physics and Center for Quantum Materials,
University of Toronto, Toronto, Ontario M5S 1A7, Canada.}
\author{Yong Baek Kim}
\affiliation{Department of Physics and Center for Quantum Materials,
University of Toronto, Toronto, Ontario M5S 1A7, Canada.}
\affiliation{School of Physics, Korea Institute for Advanced Study, Seoul 130-722, Korea.}

\begin{abstract}
Motivated by recent experiments on $\beta-$Li$_2$IrO$_3$, we study the phase diagram of the Heisenberg-Kitaev model on a three dimensional lattice of tri-coordinated Ir$^{4+}$, dubbed the hyperhoneycomb lattice by Takagi {\it et. al}. The lattice geometry of this material, along with Ir$^{4+}$ ions carrying $J_{\rm eff}=1/2$ moments, suggests that the Heisenberg-Kitaev model may effectively capture the low energy spin-physics of the system in the strong-coupling limit. Using a combination of semiclassical analysis, exact solution and slave-fermion mean field theory, we find, in addition to the spin-liquid, {\it four} different magnetically ordered phases depending on the parameter regime. All four magnetic phases--the N\'{e}el, the polarized ferromagnet, the {\it skew-stripy} and the {\it skew-zig-zag}, have collinear spin ordering. The three dimensional Z$_2$ spin liquid, which extends over an extended parameter regime around the exactly solvable Kitaev point, has a gapless Majorana mode with a deformed Fermi-circle (co-dimensions, $d_c=2$). We discuss the effect of the magnetic field and finite temperature on different phases that may be relevant for future experiments.
\end{abstract}
\date{\today}
\maketitle
\section{Introduction}

Recent studies show that 5d transition metal (Ir=iridium, Os=osmium) oxides,\cite{PhysRevB.82.064412,PhysRevLett.108.127203,PhysRevLett.110.076402,PhysRevB.83.220403,PhysRevLett.108.127204,JPSJ.70.2880,matsuhira2007metal,qi2012strong,nakatsuji2006metallic} with large spin-orbit coupling, are promising candidates for realizing a number of previously unknown electronic phases of matter\cite{RevModPhys.82.3045,RevModPhys.83.1057,doi:10.1146/annurev-conmatphys-062910-140432,pesin2010mott,yang2010topological,witczak2013correlated} as well as providing concrete material systems that may harbour some of the so far theoretically studied novel quantum phases of electrons.\cite{wan2011topological,Kitaev20062,nussinov2013compass} To this latter category belongs the now well-known Kitaev model.\cite{Kitaev20062} Originally proposed on a honeycomb lattice, the Kitaev model is an  exactly solvable spin-1/2 Hamiltonian that has a quantum spin-liquid ground state. Subsequent studies found similar exactly solvable spin models on several other two and three dimensional lattices.\cite{PhysRevB.79.024426,Si2008428,PhysRevLett.99.247203,PhysRevB.76.180404,PhysRevB.83.180412,nussinov2008bond}

In an interesting work by Jackeli {\it et al.}\cite{PhysRevLett.102.017205}, it was pointed out that in presence of strong SO coupling, spin Hamiltonians of the kind proposed by Kitaev ({\it quantum compass models}) can be realized in certain 5d transition metal oxide Mott insulators with coordination number $z=3$. While the almost simultaneous discovery of two honeycomb iridium oxide Mott insulators (Na$_2$IrO$_3$\cite{PhysRevB.82.064412} and Li$_2$IrO$_3$\cite{PhysRevLett.108.127203}) have led to a thorough investigation of these Hamiltonians on the honeycomb lattice, there are other tri-coordinated lattices in both two and three spatial dimensions, where similar physics may become relevant in the context of materials. 

\begin{figure}
\centering
\setlength\fboxsep{0pt}
\setlength\fboxrule{0.0pt}
\fbox{\includegraphics[scale=0.1,clip=true,trim=50 250 0 0]{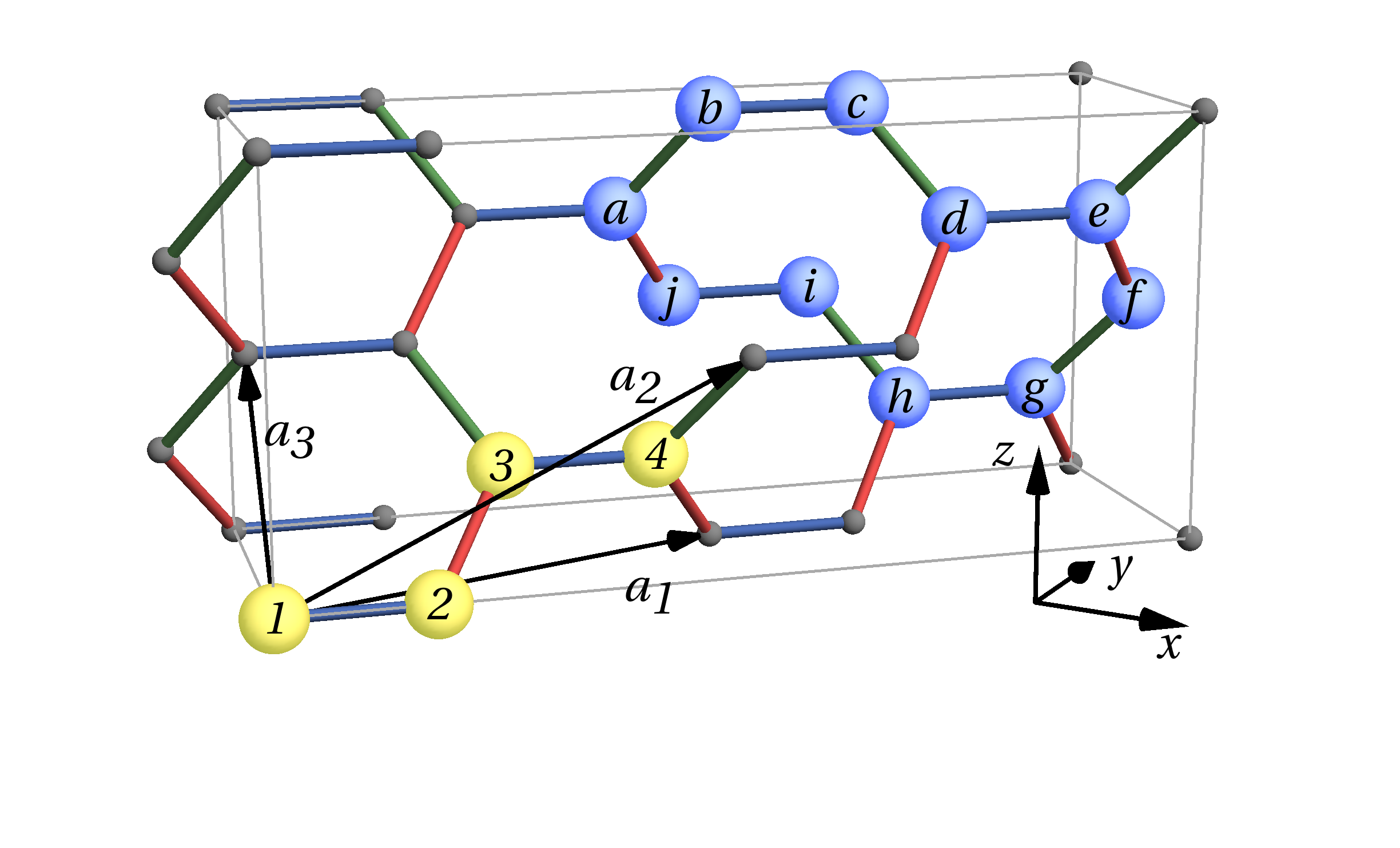}}
\caption{(color online) The tri-coordinated orthorhombic lattice. The orthorhombic unit cell is outlined in gray. The primitive unit cell contains {\it four} Ir atoms colored yellow and are labeled from 1 to 4. The ten blue sites show the smallest closed loop on this lattice.  These sites are labeled from \textit{a} to \textit{j}. All the other Ir atoms are colored gray. The primitive vectors for the 4-site unit cell are given by ${\bf a}_{i}$. For the Kitaev interactions, the red bonds refer to $S^xS^x$, the green to $S^yS^y$, and the blue to $S^zS^z$ interactions respectively.  The orientation of the global $x,y,z$ coordinates are shown in the bottom right.}
\label{fig_lattice}
\end{figure}

In this work, we study such a three dimensional Ir based Mott insulator where the magnetism may be correctly described by a generalized {\it quantum compass} Hamiltonian. Our work is directly motivated by the recent experiments by H. Takagi {\it et al.}\cite{2013_takagi} on $\beta$-Li$_2$IrO$_3$. In this material, the Ir$^{4+}$ ions, carrying $J_{\rm eff}=1/2$ moments, sit on a three dimensional network that has been dubbed a hyperhoneycomb lattice (face-centred-orthorhombic lattice with a 4-site unit cell) by Takagi {\it et al.}\cite{2013_takagi} (Fig. \ref{fig_lattice}). Since each Ir site has three Ir neighbours and is surrounded by an oxygen octahedron (see below), we find that a spin-1/2 {\it quantum compass} model captures the low energy spin physics of this system in the strong coupling limit (with localized moments). 

This is particularly interesting  and our study shows that on the present lattice the above Hamiltonian allows, apart from {\it four} magnetically ordered phases, a quantum spin liquid phase over an extended part of the phase diagram. This spin liquid is adiabatically connected to the exactly solvable ground state of the Kitaev model. We use a combination of semiclassical analysis (Luttinger-Tisza approximation with zero point corrections from spin-waves), exact solution and slave-fermion mean field theory to find the details of the phase diagram over the entire parameter regime. We find that all the magnetic phases, namely, the N\'{e}el, the polarized ferromagnet, the {\it skew-stripy} (Fig. \ref{fig_stripy}) and the {\it skew-zig-zag} (Fig. \ref{fig_zz}), have collinear spin ordering. The last two phases (see below) have interesting similarities and important differences with their two dimensional counterparts obtained on the honeycomb lattice.\cite{PhysRevLett.102.017205,PhysRevLett.105.027204} The spin liquid, on the other hand, is a three dimensional Z$_2$ spin liquid, with a gapless Majorana spinon mode. The Majorana spinon has gapless line nodes (a Fermi-circle) which is a Fermi-surface with co-dimension, $d_c=2$.  It is therefore interesting to ask if any of the above phases are relevant in explaining the magnetic properties of $\beta$-Li$_2$IrO$_3$ or similar compounds. 

The rest of the paper is organized as follows. We start, in Section \ref{sec_ham}, by discussing the details of the hyperhoneycomb lattice and the relevance of the Heisenberg-Kitaev spin Hamiltonian for $\beta$-Li$_2$IrO$_3$. In Section \ref{sec_special}, we discuss the special points in the phase diagram where the Hamiltonian becomes particularly tractable. These include the $K=0$ point where the N\'{e}el state is the classical ground state. Similarly, for $K=2J$ and $J=0$, the Hamiltonian becomes exactly solvable. While the former gives a magnetically ordered ground state, the latter is a gapless three dimensional Z$_2$ spin liquid which is the ground state for the exactly solvable Kitaev model. Following this, we investigate the general phase diagram of the Heisenberg-Kitaev model on the hyperhoneycomb lattice in the classical limit in Section \ref{sec_lt}. We discuss the {\it four} different kinds of magnetic orders in different parameter regimes-- the N\'{e}el, the {\it skew-stripy}, the {\it skew zig-zag} and the ferromagnet.  Specializing to the skew-stripy phase, we find that although stripy orders in various directions have the same energy classically, the zero-point energy corrections coming from the spin-waves split this accidental degeneracy and favours the so-called $z$-skew-stripy phase. We study the spin-wave spectrum of this phase. Following this, we study the spin liquid regime in Section \ref{sec_sf}. Since the Heisenberg term is a short range four-fermion interaction in terms of the Majorana fermionic spinons (which form the quasi-particles at the exactly solvable Kitaev limit), tree-level scaling suggests that it is irrelevant at the Kitaev fixed point and hence a finite value of $J$ is required to cause a phase transition out of the spin liquid. We study the effect of the perturbation as well as the transition using a slave-fermion mean field theory. We show how the mean-field theory connects to the exact solution at the Kitaev point. Our analysis gives a first order transition from the spin liquid to the skew-stripy phase. Response to finite temperature and magnetic field are briefly discussed in Section \ref{sec_magt} for both the skew-stripy phase and the spin liquid phase. Finally we summarize our results in Section \ref{sec_discuss}. Details of various calculations are given in different appendices.

\section{The lattice and the Hamiltonian}
\label{sec_ham}

The geometry of the compound suggests that each Ir$^{4+}$ ion sits inside an oxygen octahedron. In such an environment, the cubic crystal field ($10Dq\sim 3$ eV) and large atomic SO coupling ($\lambda\sim 500$ meV) in Ir split the 5d orbitals into lower $J_{\rm eff}=3/2$ and the upper $J_{\rm eff}=1/2$ atomic orbitals. The five electrons of Ir$^{4+}$ completely fills the quadruplet, while leaving the doublet half filled. Thus the low energy magnetism is expected to be described by the latter orbitals which form a $J_{\rm eff}=1/2$ pseudo-spin at each Ir$^{4+}$ site.\cite{Kim06032009}  

The network of Ir$^{4+}$ ions then form a tri-coordinated network as shown in Fig. \ref{fig_lattice} (further details are discussed in Appendix \ref{appen_hyperhoneycomb}.). This Ir$^{4+}$ ion network is topologically equivalent (not shown) to a decorated diamond lattice (where each site of the diamond lattice is split into two) or a depleted cubic lattice\cite{PhysRevB.79.024426}. The neighbouring oxygen octahedra share edges with Ir-O-Ir and Ir-Ir-Ir angles being 90$^\circ$ and 120$^\circ$ respectively in the ideal structure. 

Before moving on, we briefly discuss the symmetries of the hyperhoneycomb lattice for future use. There are three types of symmetry operations in the hyperhoneycomb:
\begin{itemize}
\item Inversion at the bond center of Ir$_2$-Ir$_3$ and Ir$_1$-Ir$_4$ (green and red bonds in Fig. \ref{fig_lattice});

\item Three orthogonal $C_2$ axes at the bond center of Ir$_1$-Ir$_2$ and Ir$_3$-Ir$_4$ (blue bonds).  These axes are parallel to the face-center-orthorombic lattice vectors $\mathbf{a}$, $\mathbf{b}$, and $\mathbf{c}$ (see Appendix \ref{appen_hyperhoneycomb} for definition of ${\bf a, b}$ and ${\bf c}$).  Ir$_2$-Ir$_3$ and Ir$_1$-Ir$_4$ bonds are interchanged via these $C_2$ axes;

\item Glide planes with translation $\mathbf{a}_i/2$ interchanges Ir$_1$-Ir$_2$ and Ir$_3$-Ir$_4$.
\end{itemize}

A strong coupling calculation using a hopping Hamiltonian with Slater-Koster parameters (similar to Jackeli {\it et al.}\cite{PhysRevLett.102.017205}), in presence of Hund's coupling and onsite Coulomb repulsion, results in the Heisenberg-Kitaev spin Hamiltonian, to the leading order.:
\begin{align}
\mathcal{H}_{\rm HK}=J\sum_{\langle ij\rangle}{\bf S}_i\cdot{\bf S}_j-K\sum_{\langle ij\rangle,\alpha-links} S^\alpha_iS^\alpha_j .
\label{eq_HK_ham}
\end{align}
The first term represents the usual Heisenberg interactions while the last term is the Kitaev exchange. The $\sum_{\langle ij\rangle, \alpha-links}$ is a standard notation used in a Kitaev model which means that on a lattice with coordination number $z=3$, there are three kinds of spin exchanges. This is depicted for the lattice of our interest in Fig. \ref{fig_lattice}. 

On occasion, we also use the one variable parametrization in terms of $\alpha$ which has been used in the honeycomb case. The relation between $J, K$ and $\alpha$ is given by:
\begin{align}
J=1-\alpha,~~~~K=2\alpha .
\end{align}

Further perturbations to $\mathcal{H}_{\rm HK}$ on the hyperhoneycomb lattice may include further neighbour exchanges as well as Dzyaloshinski-Moriya (DM) interactions. The inversion center ensures that the DM vector vanishes for red/green ($x/y$) bonds (Ir$_2$-Ir$_3$/Ir$_1$-Ir$_4$) for a $J_{\text{eff}}=1/2$ pseudo-spin model.  The $C_2$ axes ensure that the Kitaev term is along $\mathbf{c}$ and the DM vector points along the bonds for blue ($z$) bonds (Ir$_1$-Ir$_2$ and Ir$_3$-Ir$_4$), i.e. along $\mathbf{a}$ (see Appendix \ref{appen_hyperhoneycomb} for definition of ${\bf a, b}$ and ${\bf c}$).  However, we find that the magnitude of the DM vector for the nearest neighbours is zero when we consider the shortest exchange paths that pass through only oxygens sites between two given Ir-sites connected through $z$-bond. Longer  exchange paths that involve intermediate Ir$^{4+}$ ions as well as oxygens, in principle, can generate a DM term along $z$-bond, but they are expected to be weak and hence we neglect them in the present calculation.

For the rest of this work, we assume that the further neighbour terms are small and the essential features of the strong coupling limit (with localized magnetic moments) of the real material is captured by $\mathcal{H}_{\rm HK}$.


\section{The special limits of the the Heisenberg-Kitaev Hamiltonian}
\label{sec_special}

We start by discussing the special limits of the $\mathcal{H}_{\rm HK}$ (Eq. \ref{eq_HK_ham}) that gives us important insight into the phase diagram. These special points are given.  (A) $K=0 (\alpha=0)$ limit which is the pure nearest neighbour Heisenberg antiferromagnet on the hyperhoneycomb lattice. (B) $K=2J (\alpha=1/2)$.  When using a 4-sub-lattice rotation, one can map the Hamiltonian to a nearest neighbour ferromagnet on the given lattice. Hence, this point is exactly solvable.  (C) $J=0 (\alpha=1)$, which is the limit for the pure Kitaev model, which on this lattice is exactly solvable. Below we discuss these three special points in detail. 

\subsection{$K=0$: N\'{e}el order}

\begin{figure}
\centering
\setlength\fboxsep{0pt}
\setlength\fboxrule{0.0pt}
\fbox{\includegraphics[scale=.1,clip=true,trim=50 250 0 0]{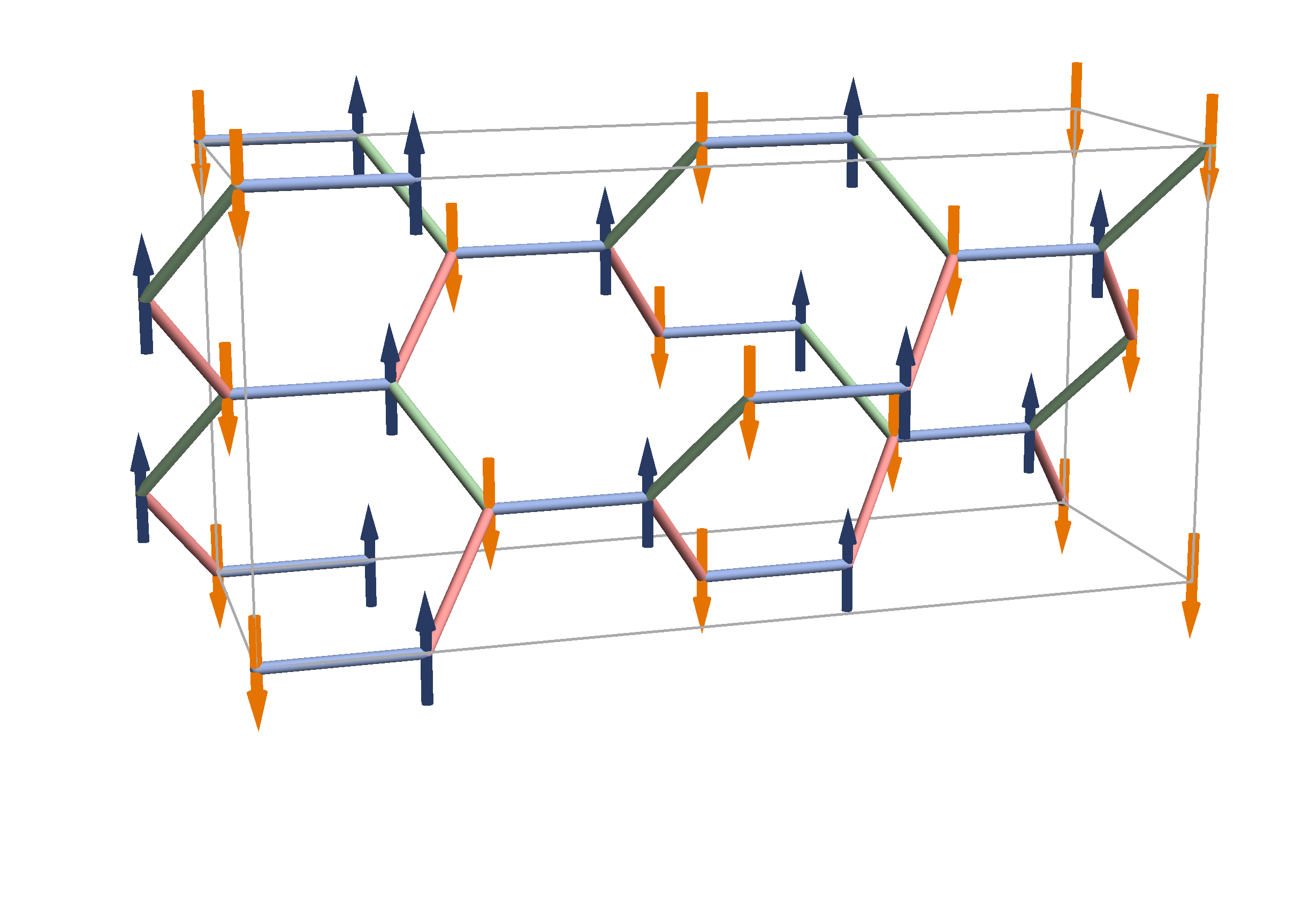}}
\caption{(color online) The N\'{e}el phase. This is the classical ground state for $K=0$.}
\label{fig_neel}
\end{figure}

This is the limit of the pure nearest neighbour antiferromagnetic Heisenberg model. As pointed out above, the present Ir$^{4+}$ network is similar topologically to a decorated diamond lattice where each site of the diamond lattice is split into two. The nearest neighbour Heisenberg antiferromagnet on this network is {\it not} frustrated at the classical level. The magnetic order is shown in Fig. \ref{fig_neel}. This classical order, in three spatial dimensions, is expected to be robust to quantum fluctuations.

\subsection{$K=2J$: Skew-Stripy Order}

\begin{figure}
\centering
\setlength\fboxsep{0pt}
\setlength\fboxrule{0.0pt}
\fbox{\includegraphics[scale=.1,clip=true,trim=50 110 0 -100]{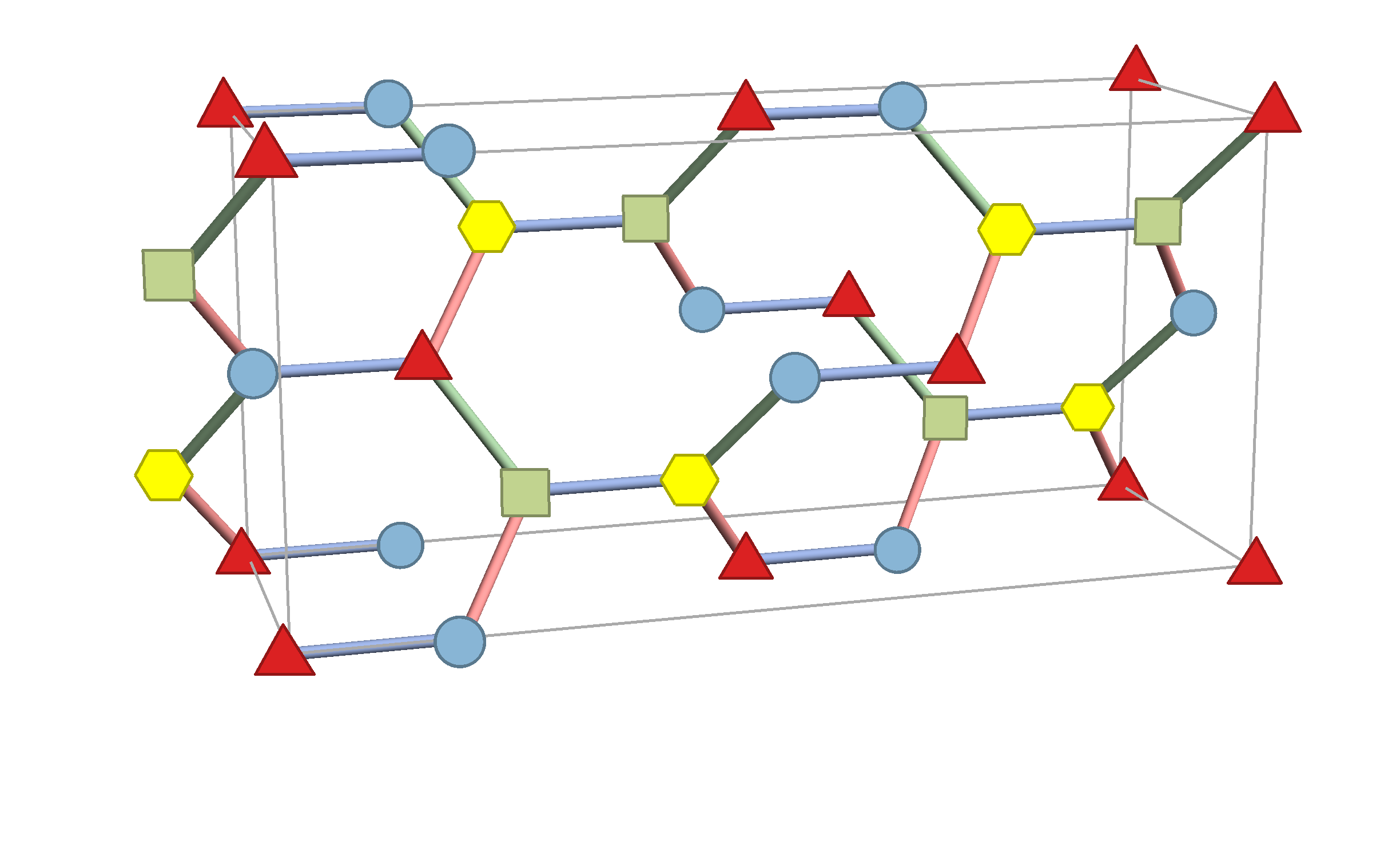}}
\caption{(color online) The equivalent of the four-sublattice rotation defined by G. Khaliullin \cite{khaliullin2005orbital} and later by Chaloupka {\it et al.} \cite{PhysRevLett.105.027204} for the hyperhoneycomb lattice. The spins at the sites denoted by blue circles are left unrotated, the spins at the sites denoted by red triangles are rotated by 180 degrees about the $z$-axis, the spins at the sites denoted by yellow hexagons are rotated by 180 degrees about the $y$-axis, and the spins at the sites denoted by green squares are rotated by 180 degrees about the $x$-axis.}
\label{fig_rot}
\end{figure}

\begin{figure}
\centering
\setlength\fboxsep{0pt}
\setlength\fboxrule{0.0pt}
\fbox{\includegraphics[scale=.1,clip=true,trim=50 250 0 0]{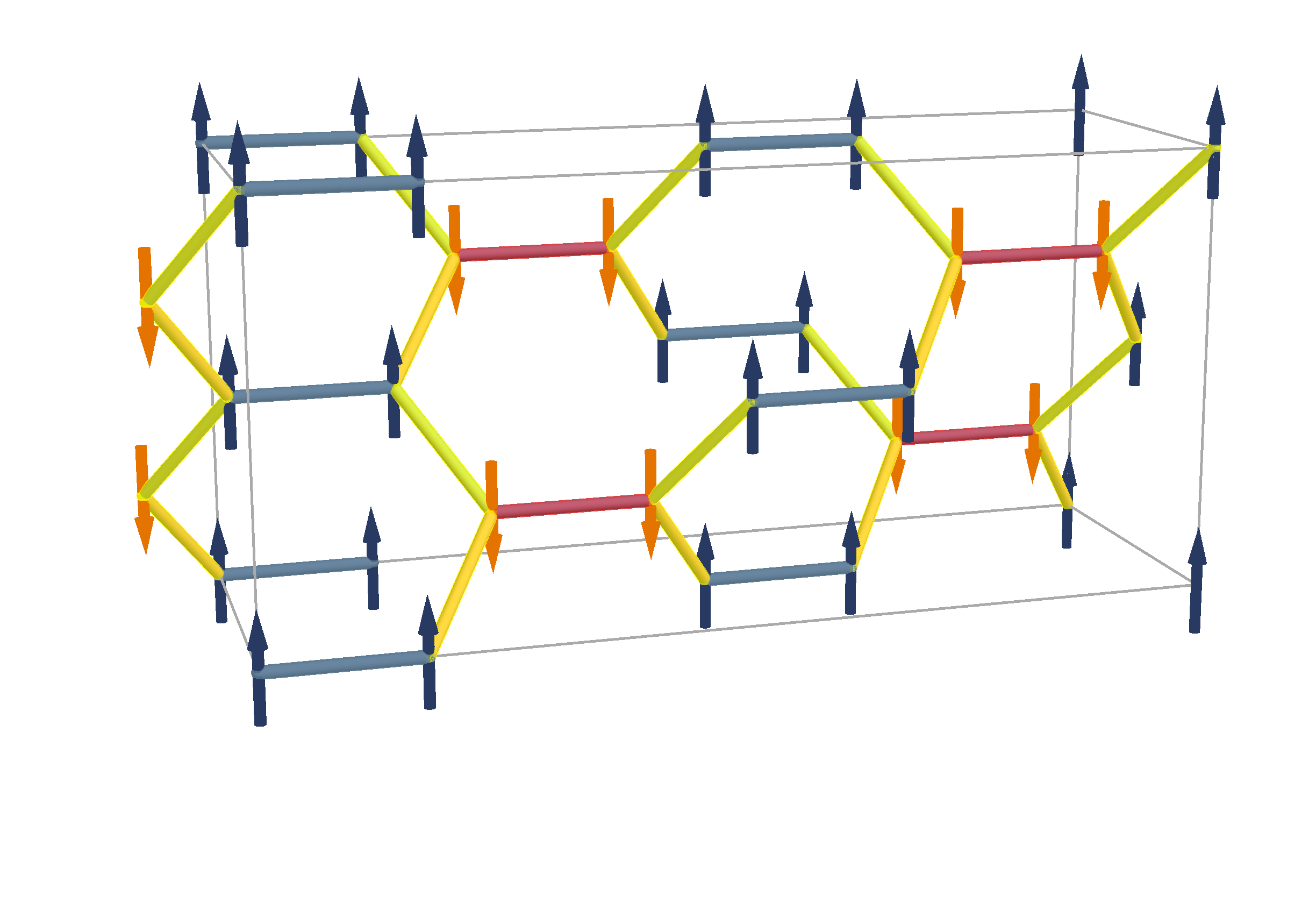}}
\caption{(color online) The {\it skew-stripy} phase with ordering in $S^z$. This is the exact solution to the model at the point $K=2J$. The antiferromagnetic chains run along the $x-y$ bonds (shaded in yellow) which form almost skew lines. The ferromagnetic $z$-bonds form a stripy order (shaded in red and blue).}
\label{fig_stripy}
\end{figure}

Similar to the case of honeycomb lattice, we can perform a site dependent rotation,\cite{khaliullin2005orbital,PhysRevLett.105.027204} defined by Fig. \ref{fig_rot}. In the rotated basis, the parameters $J$ and $K$ map as $J \rightarrow -J$ and $K \rightarrow K-2J$.\cite{PhysRevB.86.224417} Upon performing this transformation, at the special point $K=2J$ we find that the Kitaev term vanishes exactly and the model describes a fully polarized ferromagnet in the rotated basis. 

The quantum ferromagnet can be exactly solved and this exact solution, when re-rotated back to the original spins, maps to a three dimensional collinear magnetic order which we call the {\it skew-stripy} state (shown in Fig. \ref{fig_stripy}). At this point, the ferromagnet can choose its axis of quantization in any direction which corresponds to different skew stripy ordering. However, as we shall see later, only three collinear states are selected by quantum fluctuations away from this point. In these three  states, the spins are aligned along $x,y$ or $z$ axes. In Fig. \ref{fig_stripy}, we have drawn the ordering in $S^z$ where the antiferromagnetically ordered chains run along the $x-y$ bonds which are then coupled ferromagnetically with the $z$-bonds. The $x-y$ bonds form chains that, in three dimensions, by themselves are ``skew" to one another as shown in Fig. \ref{fig_stripy} and the ferromagnetic z-bonds joining such chains alternate from having up spin to down spins. Hence we call this the {\it skew-stripy} phase. The $x$ and the $y$ phases similarly have ferromagnetic $x$ or $y$ bonds coupling skew chains running along the $y-z$ and $x-z$ bonds, respectively. We would like to re-emphasize that the word {\it skew} indicates that this is essentially a three dimensional magnetic order as opposed to a stacked up two dimensional spin order. At this special point there is a continuous ``\textit{SU(2)}" spin rotation symmetry that ensures that all the three skew-stripy phases described above have the same energy.

\begin{figure*}
\centering
\setlength\fboxsep{0pt}
\setlength\fboxrule{0.0pt}
\fbox{\includegraphics[scale=.1,clip=true,trim=0 0 0 0]{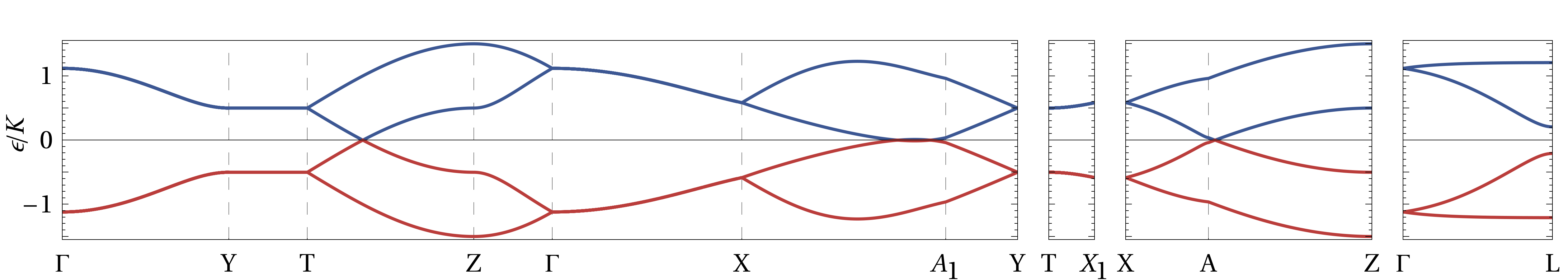}}
\caption{The spectrum of the dispersing Majorana fermion in the pure Kitaev model on the hyperhoneycomb lattice along paths of high symmetry in the first Brillouin zone (The first Brillouin zone and the paths are shown in Appendix \ref{appen_hyperhoneycomb})}
\label{fig_hyper_spectrum}
\end{figure*}

It is however worthwhile to note that there is a crucial difference from the honeycomb case away from this special point. In the honeycomb lattice  a two dimensional stripy phase is obtained for the Heisenberg-Kitaev model at the same parameter value. There, a $C_3$ symmetry of the lattice along with concomitant rotation of the spins which is a symmetry of the $\mathcal{H}_{\rm HK}$ Hamiltonian on the honeycomb lattice ensures that the three stripy ordered phases have the same energy even away from this special point where there is no \textit{SU(2)} symmetry. However on the hyperhoneycomb lattice, there is only a $C_2$ symmetry between the $x$ and the $y$ bonds, while the $z$ bonds are not related by any symmetry. So there is no \textit{a priori} reason for the $S_z$ ordered skew-stripy phase to have the same energy as the other two. Indeed we find that, away from this point ($K=2J$), although the classical energies of the three states remain the same, quantum corrections coming from the spin-wave fluctuations lift this accidental classical degeneracy.

\subsection{$J=0$ : The Kitaev Spin Liquid}

This is the pure Kitaev limit. Mandal {\it et. al} \cite{PhysRevB.79.024426} showed that the pure Kitaev model on the deleted cubic lattice which is topologically similar to the hyperhoneycomb lattice can be exactly solved using methods originally employed by Kitaev.\cite{Kitaev20062} 

The exact solution, as in the honeycomb case, is rendered by the three-fold coordination and consequent presence of an infinite number of conserved quantities. Using the usual Majorana fermion decomposition of the spins 
 
\begin{align}
S_i^\alpha=\frac{1}{2}ib_i^\alpha c
\end{align}
we find that the Hamiltonian (Eq. \ref{eq_HK_ham}) in this limit is given by:
\begin{align}
\mathcal{H}_{\rm K}=\frac{i}{2}\sum_{\alpha-{\rm links}} u^{\alpha}_{ij}c_ic_j ~~~~~~~~~~~({\rm where}~~u_{ij}^\alpha=ib_i^\alpha b_j^\alpha),
\label{eq_hop}
\end{align}
where we have put the overall scale $K/4=1$ (the quarter comes from the fact that we have spin-1/2). The $\left\{b_i^x,b_i^y,b^z_i,c\right\}$ are the four Majorana fermions that mutually anticommute. 

The infinite number of conserved quantities are given by the $Z_2$-link variables $u^\alpha_{ij}$ that commute with each other and with the Hamiltonian (Eq. \ref{eq_hop}). The Z$_2$-fluxes generated by $u^\alpha_{ij}$ over the 10 site loop (the blue sites in Fig. \ref{fig_lattice}) are given by\cite{PhysRevB.79.024426}
\begin{align}
\mathcal{W}_P=\prod_{\rm loop}u^\alpha_{ij} .
\label{eq_flux}
\end{align}
Since these fluxes commute with the Hamiltonian, by construction, they do not have any dynamics and hence the problem can be solved independently for different flux sectors. This separation of the Majorana sector and the flux sector, the latter being good quantum numbers, lies at the heart of the exact solution of the Kitaev models on different lattices.\cite{Kitaev20062}

The problem then reduces to Majorana fermions hopping in the background of frozen fluxes on the hyperhoneycomb lattice. Similar issues have been studied by various people on other lattices. E. Lieb \cite{PhysRevLett.73.2158} proved that, on certain bipartite lattice that contain mirror planes that bisect the lattice links, the lowest energy is obtained when planar plaquettes containing 2(mod 4) sites have zero-flux through them, while plaquettes having 0 (mod 4) sites have $\pi$-flux through them.  Unfortunately, unlike the 2D-honeycomb lattice, we cannot prove Lieb's theorem for the present lattice\cite{PhysRevB.79.024426} because of the absence of suitable mirror planes. In absence of such theorems, Mandal {\it et al.}\cite{PhysRevB.79.024426} resorted to numerical diagonalization of the fermion hopping Hamiltonian (Eq. \ref{eq_hop}) over large system sizes for several flux configurations and found that the zero-flux sector has the lowest energy. Thus it is expected that the zero flux sector corresponds
to the ground state in our case as well. We can then specialize to the zero-flux sector choosing a gauge where $u^\alpha_{ij}=+1$ (for particular configurations of $\langle ij\rangle$ as shown in Appendix \ref{appen_zeroflux}) to get

\begin{align}
\mathcal{H}_{\rm K}^{0-{\rm flux}}=\frac{i}{2}\sum_{ij}c_ic_j .
\label{eq_zerohop}
\end{align}

This Hamiltonian can then be diagonalized by Fourier transformation, taking the unit cell as given in Fig. \ref{fig_lattice} (the lattice vectors are given in Appendix \ref{appen_hyperhoneycomb}). We get

\begin{align}
\mathcal{H}_{\rm K}^{0-{\rm flux}}=\sum_{\bf k}\Psi_{-\bf k}^TH_{\bf k}\Psi_{\bf k}
\end{align}
where $ \Psi_{\bf k}^T=\left(c_{1,\bf k},c_{2,\bf k},c_{3,\bf k},c_{4,\bf k}\right)$ and

\begin{align}
H_{\bf k}=\frac{i}{4}\left(\begin{array}{cccc}
0 & 1 & 0 & \mathcal{A}_{\bf k}\\
-1 & 0 & \mathcal{B}_{\bf k} & 0\\
0 &-\mathcal{B}_{\bf k}^* & 0 & 1\\
-\mathcal{A}_{\bf k}^* & 0 & -1 & 0\\
\end{array}\right)
\end{align}
where,
\begin{align}
\mathcal{A}_{\bf k}=e^{-i{\bf k\cdot a_1}}+e^{-i{\bf k\cdot a_2}};~~~~~~~~\mathcal{B}_{\bf k}=1+e^{-i{\bf k\cdot a_3}}
\end{align}
The spectrum is given by:
\begin{align}
&\mathscr{E}_{\bf k}=\pm\frac{1}{2\sqrt{2}}\left[(2+|\mathcal{A}_{\bf k}|^2+|\mathcal{B}_{\bf k}|^2)\right.\nonumber \\
&\pm\left.\sqrt{\left[2+|\mathcal{A}_{\bf k}|^2+|\mathcal{B}_{\bf k}|^2\right]^2-4\left[1+|\mathcal{A}_{\bf k}|^2|\mathcal{B}_{\bf k}|^2+2\Re\left[\mathcal{A}_{\bf k}\mathcal{B}_{\bf k}^*\right]\right]}\right]^{1/2}
\label{eq_disp}
\end{align}

\begin{figure}
\centering
\setlength\fboxsep{0pt}
\setlength\fboxrule{0.0pt}
\fbox{\includegraphics[scale=0.1,clip=true,trim=100 700 -100 400]{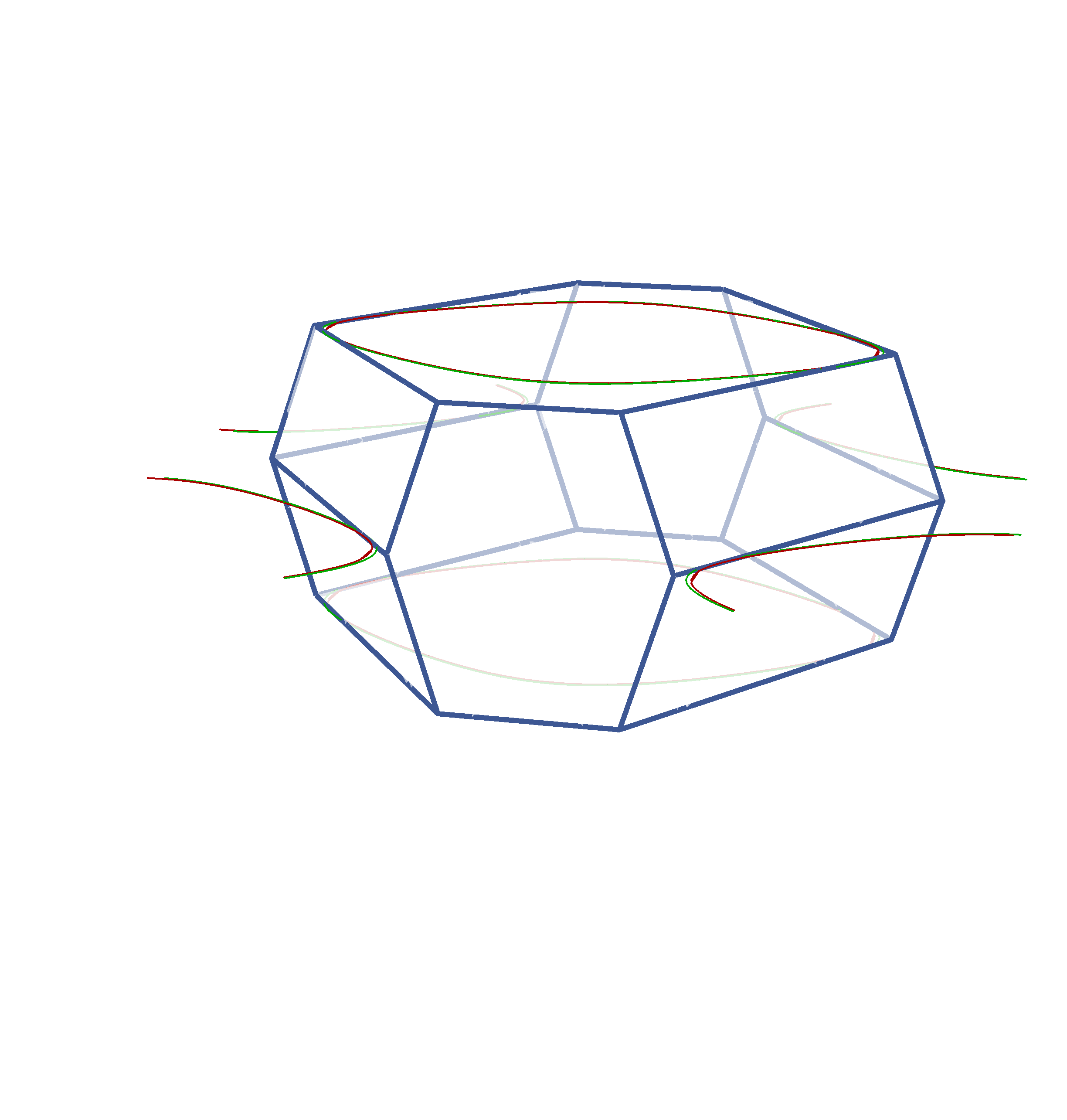}}
\caption{(color online) The green curve indicates the Fermi surface at $J=0$. This occurs on the boundary of the first Brillouin zone.  The red curve indicates the Fermi surface at $K/J=8 (\alpha=0.2)$ as computed within mean-field theory.  The two Fermi surfaces almost coincide with minute differences.  (The Fermi surfaces of neighboring cells have been appended to aid visualization.)}
\label{fig_fermi}
\end{figure}

The spectrum for the dispersing Majorana fermion, $c$, along the high symmetry lines within the first Brillouin zone is given in Fig. \ref{fig_hyper_spectrum}. The lower two bands are occupied while the zero energy surface describe the contour of the gapless excitation. We find a Fermi surface of co-dimension {\it two}, {\it i.e.} line nodes.  From Eq. \ref{eq_disp}, it is easy to see that this is given by the zeros of the term $1+|\mathcal{A}_{\bf k}|^2|\mathcal{B}_{\bf k}|^2+2\Re\left[\mathcal{A}_{\bf k}\mathcal{B}_{\bf k}^*\right]$, which can be rewritten as $|1+\mathcal{A}_{\bf k}\mathcal{B}_{\bf k}^*|^2$. A straightforward manipulation of this expression reveals that this can occur only when $k_x + k_y \equiv 0$ (${\rm mod}~ \frac{\pi}{3}$), and $\cos{(ky-kx)} + \cos{(2k_z)} = \pm \frac{1}{2}$ (with the sign determined by $k_x + k_y$). This determines the exact location of this Fermi surface which is shown in Fig. \ref{fig_fermi}. The line nodes occur in the zone-boundary as shown. The presence of these extended gapless modes have important finite temperature consequences as we discuss later.

The Majorana-spinon representation enlarges the dimension of the Hilbert space from two to four. Therefore, the physical  spin wave function is obtained by projecting the spinon wave function back to the physical Hilbert space.\cite{Kitaev20062,PhysRevB.79.024426}
\begin{align}
\vert\Psi_{\rm spin}\rangle=\mathcal{P}\vert\Psi_{\rm spinon}\rangle
\end{align}
where the projection operator, $\mathcal{P}$, is given by
\begin{align}
\mathcal{P}=\prod_i\left(\frac{1+\mathcal{D}_i}{2}\right)
\end{align}
where,
\begin{align}
\mathcal{D}_i=b^x_ib^y_ib^z_ic_i
\end{align}
and in the physical Hilbert space, the spinon wave function satisfies $(\prod_iD_i)|\Psi_{\rm spinon}\rangle=|\Psi_{\rm spinon}\rangle$.\cite{Kitaev20062} The gauge invariant Z$_2$ flux operator in Eq. \ref{eq_flux} can be written in terms of the spin variables as
\begin{align}
\mathcal{W}_p=2^{10}S^x_bS^x_cS^x_dS^y_eS^z_fS^x_gS^x_hS^x_iS^y_jS^z_a
\label{eq_spinflux}
\end{align}
where the numberings refer to sites as shown in Fig. \ref{fig_lattice}. The rule for writing the expression of $\mathcal{W}_p$ in terms of the spins is similar to the original Kitaev model\cite{Kitaev20062}---for the site $i$, if the bonds participating in the loop are of $x$ and $y$ types (note they cannot be of the same type by construction), then $\mathcal{W}_p$ contains the third component of the spin, {\it i.e.} $S_i^z$. The flux operator is constructed by repeating this procedure. There are {\it four} different kinds of 10-loop plaquettes\cite{PhysRevB.79.024426}.

This ends our discussion on the special limits of the Heisenberg-Kitaev Hamiltonian. Next, we shall discuss the general phase diagram first at the classical limit within Luttinger Tisza approximation and then in the quantum limit using slave-fermion mean field theory.

\section{Classical phase diagram within Luttinger-Tisza approximation and spin-wave analysis}
\label{sec_lt}

\begin{figure}
\centering
\setlength\fboxsep{0pt}
\setlength\fboxrule{0.0pt}
\fbox{\includegraphics[scale=1,clip=true,trim=0 0 0 0]{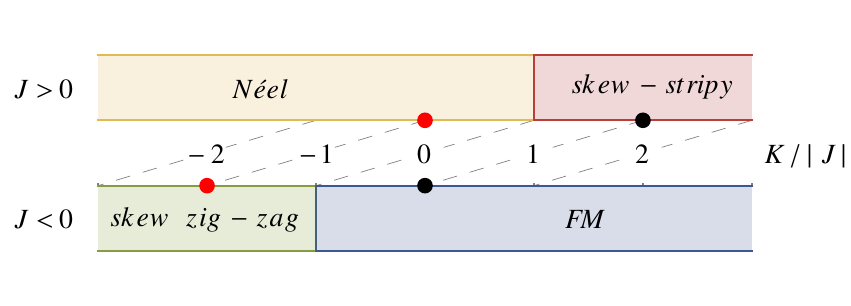}}
\caption{(color online) The classical phase diagram within Luttinger-Tisza approximation for arbitrary $J$ and $K$.  Diagonal dotted lines indicate the four-sublattice rotation mapping from $(J,K) \rightarrow (-J,K-2J)$.  Black dots indicate the exactly solvable ferromagnetic point in both the rotated and the unrotated bases.  Red dots indicate the antiferromagnetic Heisenberg point in both the rotated and the unrotated bases.  Four magnetic phases have been found, see main text for details.}
\label{fig_LT_pd}
\end{figure}

Beyond the special points as discussed above, we can study the general phase diagram of the Hamiltonian in Eq. \ref{eq_HK_ham} in the classical limit within the Luttinger-Tisza approximation\cite{PhysRev.70.954} for arbitrary $J$ and $K$.  The phase diagram is shown in Fig. \ref{fig_LT_pd}.  Four magnetic orders are found: they are the N\'{e}el, skew-zig-zag, skew-stripy, and ferromagnetic order. It is noteworthy that all the magnetically ordered phases shown here have their counterpart in the honeycomb case, though with important differences, and hence we have used a similar nomenclature. 

\begin{figure}
\centering
\setlength\fboxsep{0pt}
\setlength\fboxrule{0.0pt}
\fbox{\includegraphics[scale=.1,clip=true,trim=50 250 0 0]{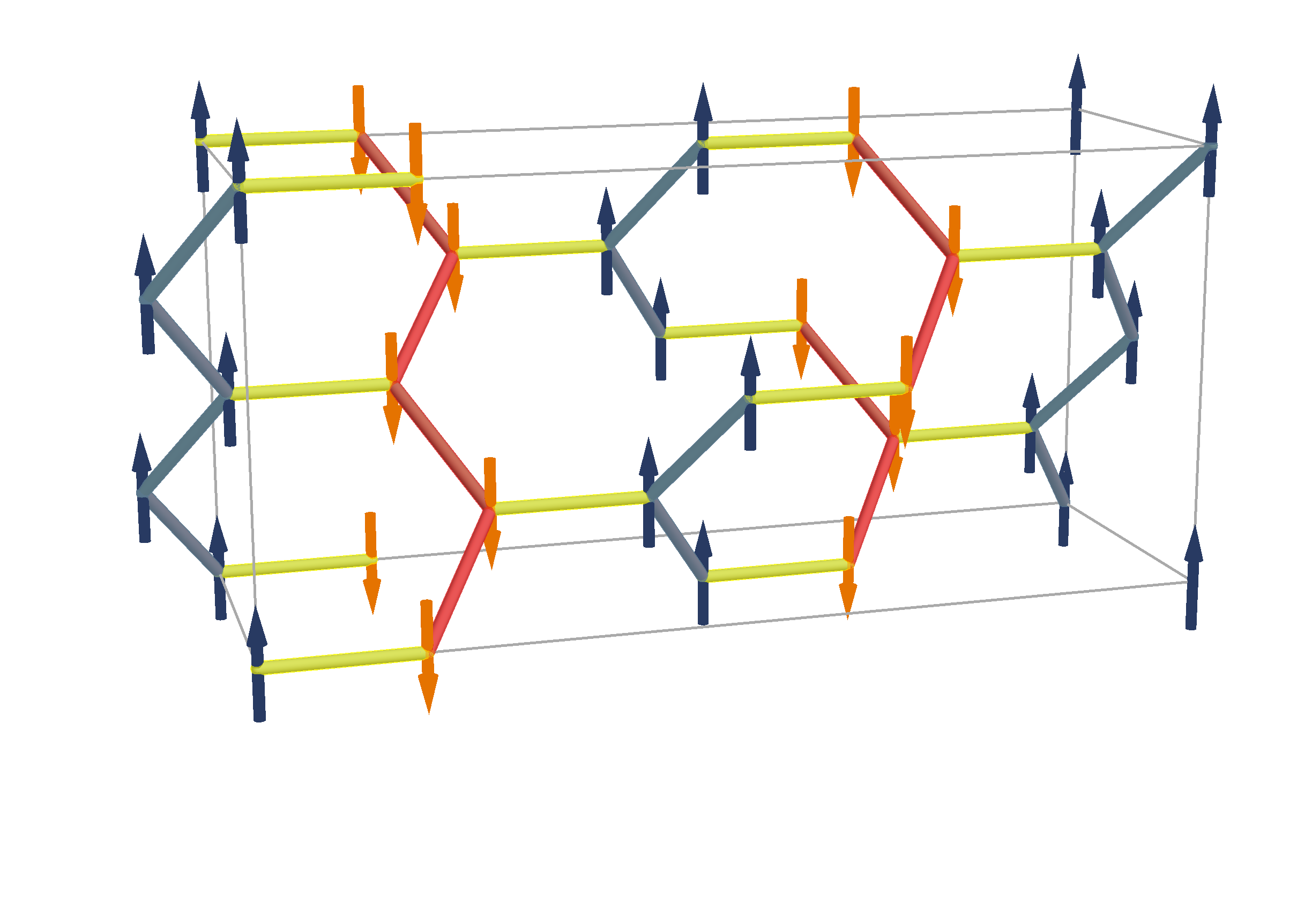}}
\caption{(color online) The skew-zig-zag phase with $S^z$ ordering. The ferromagnetic chains run along the $x-y$ bonds in a zig-zag fashion (indicated in blue and red), while the $z$-bonds are ferromagnetic (indicated in yellow).}
\label{fig_zz}
\end{figure}

Although, in the rest of this paper, we mainly concentrate on the parameter regime $J,K>0$, here we note that it is sufficient at the classical level, as shown in Fig. \ref{fig_LT_pd}, to study the $J>0$ region of the phase diagram. The $J<0$ part of the phase diagram is easily   obtained using the aforementioned four-sublattice rotation. The N\'{e}el and skew-zig-zag orders are related by the rotation, as are the ferromagnetic and stripy orders.  The skew-zig-zag order in Fig. \ref{fig_zz} is ordered in the $S^z$ direction. In contrast to the skew-stripy, this has ferromagnetically aligned chains running in along the $x-y$ bonds which are then connected antiferromagnetically along the $z$-bonds. Similar to the skew-stripy, this is also an inherently three dimensional magnetic order.

\subsection*{Spin-wave zero-point corrections about the classical solution}
\begin{figure*}
\centering
\setlength\fboxsep{0pt}
\setlength\fboxrule{0.0pt}
\fbox{\includegraphics[scale=.1,clip=true,trim=0 0 0 0]{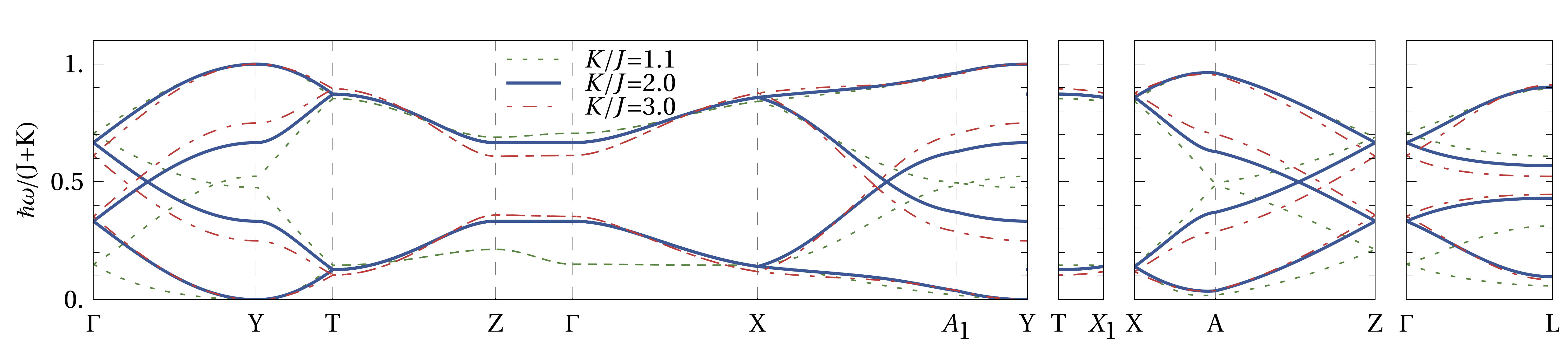}}
\caption{The spin-wave dispersion for various values of $K/J$ within the skew-stripy phase. We have chosen the ordering in $S^z$ as an example. $K/J=2 (\alpha=0.5)$ maps to the pure ferromagnetic model in the rotated basis (see text); $K/J=1.1 (\alpha=0.35)$ is near the classical boundary of the N\'{e}el and the skew-stripy order; and, $K/J=0.6$ is a general point within the skew-stripy phase.}
\label{fig_spinwaves_stripy}
\end{figure*}

As pointed out before, at the classical level the Heisenberg-Kitaev Hamiltonian has a spurious \textit{SU(2)} symmetry and because of this, the different skew-stripy ordered phases have the same classical energy. However, since this degeneracy is accidental, quantum fluctuations in the form of zero-point corrections coming from the spin-waves break the above degeneracy (see below).  We study the quadratic spin-wave theory using the Holstein-Primakoff bosons.

In Fig. \ref{fig_spinwaves_stripy}, we plot the spin-wave dispersion for different values of $K/J$ for the $z$-skew-stripy phase.  There is a gapless Goldstone mode for $K=2J$ at the zone-boundary $Y$. This is a consequence of the fact that the 4-sublattice rotation, at this point, maps the system exactly to a ferromagnet. Indeed, the mode is quadratically dispersing ($\omega\sim k^2$), as is expected for a ferromagnet. However, we find, similar to the honeycomb case, this gapless mode is present for all values of $K/J$ in the skew-stripy regime ($K>J>0$). This is due to the spurious \textit{SU(2)} symmetry at the classical level which survives even for the quadratic spin-wave theory.  However, this gapless mode is not protected by symmetry of the general Hamiltonian and higher order corrections coming from magnon-magnon interactions gaps out this mode.

\begin{figure}
\centering
\setlength\fboxsep{0pt}
\setlength\fboxrule{0.0pt}
\fbox{\includegraphics[scale=.1,clip=true,trim=0 0 0 0]{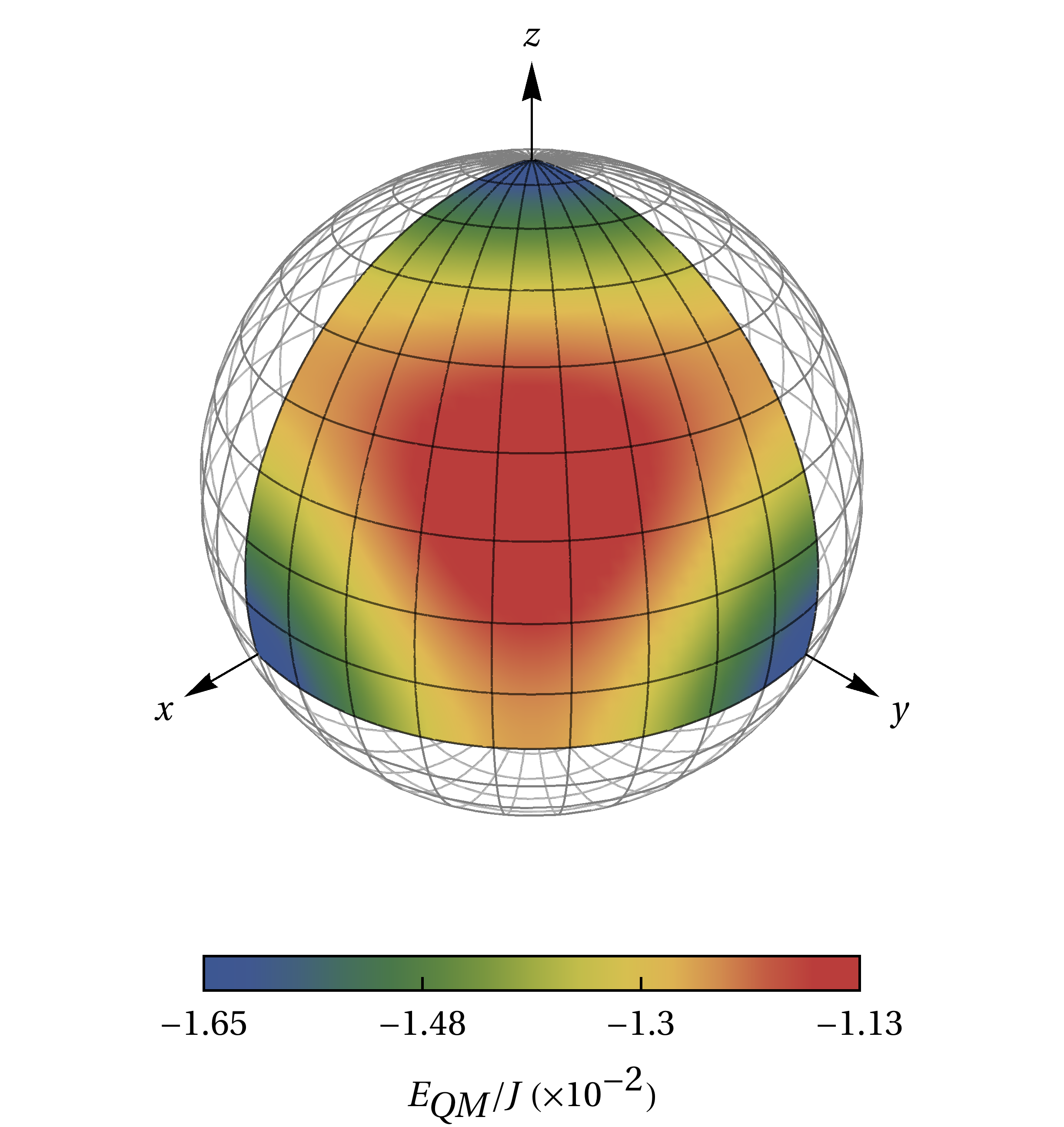}}
\caption{The zero-point energy corrections from spin-wave theory for the different skew-stripy states for $K=3J$. The different stripy states can be labeled by $(\theta,\phi)$ (see text). We find that the $x$, $y$ and the $z$ skew stripy states have lower energies (but not same, as this figure may deceptively suggest, due to lack of resolution. See Fig. \ref{fig_diff} for the difference). Due to the three $C_2$, two inversion, and time-reversal symmetries of our Hamiltonian, energy correction for other $(\theta,\phi)$ not explicitly shown is related to the plotted octant by mirror operations $\sigma_{yz}$, $\sigma_{xz}$, and $\sigma_{xy}$ in $(\theta,\phi)$ space.}
\label{fig_sphere}
\end{figure}

\begin{figure}
\centering
\setlength\fboxsep{0pt}
\setlength\fboxrule{0.0pt}
\fbox{\includegraphics[scale=.1,clip=true,trim=0 0 0 0]{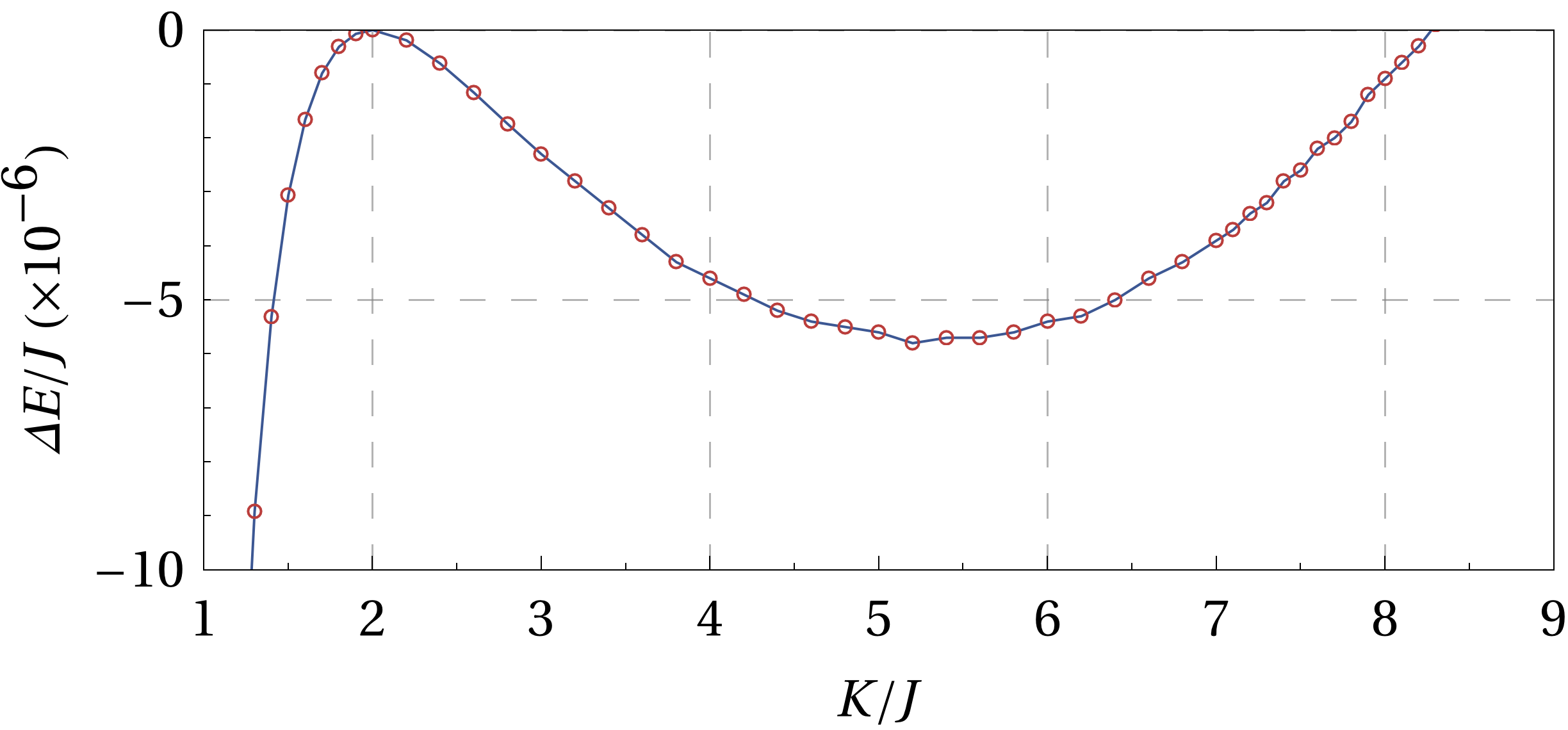}}
\caption{The zero-point energy difference between the $S^z$-ordered and $S^{x}/S^{y}$-ordered skew-stripy phases as a function of $K/J$.  Negative values indicating $S^z$-ordered phases have lower energy.  At the exactly solvable point, $K/J=2 (\alpha=0.5)$, \textit{SU(2)} symmetry is restored and hence the energy difference is zero.  Away from that point, the $S^z$-ordered phase has lower energy and hence is selected by this quantum-order-by-disorder mechanism.}
\label{fig_diff}
\end{figure}

The finite momentum of the zero-energy mode may seem counter-intuitive at first, particularly at the $K=2J$ point where the system can be rotated to a ferromagnet with uniform (${\bf q=0}$) order. However, we immediately note that this 4-sublattice rotation (Fig. \ref{fig_rot}) has an 8-site unit cell and hence has a  finite momentum (which is exactly equal to ${\bf q}=Y$) within the Brillouin zone of our 4-site unit cell (this implies that the general skew-stripy order actually has $\mathbf{q}=Y$ order within our 4-site unit cell, or, equivalently, $\mathbf{q}=0$ order within an 8-site unit cell).  Therefore, if we were to examine the spin-wave spectrum in the rotated basis, the gapless quadratic dispersion would shift to the $\Gamma$ point of the Brillouin zone.

We next calculate the zero-point energy coming from the spin-waves for different skew-stripy states. The different classical skew-stripy states can be parametrized by spherical angles $(\theta,\phi)$, where $(\theta,\phi)=(0,0)$, $(\pi/2,0)$, and $(\pi/2,\pi/2)$ are the $z$-, $x$-, and $y$-skew-stripy states respectively. One way of seeing this is that since the stripy phase is just a ferromagnet in the rotated basis, the two angles quantify the direction of quantization of this ferromagnet with $\theta$ being the polar angle (with reference to the $z$ direction) and $\phi$ being the azimuthal angle. We can then obtain the magnitude of the zero-point corrections for different states (for various values of $K/J$) as a function of $(\theta,\phi)$. As an example, the resulting ground state energy corrections for $K=3J$ as a function of $(\theta,\phi)$ is given in Fig.\ref{fig_sphere}.

The variation in energy correction as a function of $(\theta,\phi)$ signifies lifting of the accidental \textit{SU(2)} symmetry.  On the other hand, discrete symmetries mentioned in Sec. \ref{sec_ham} are manifested as mirrors planes in parameter space ($\sigma_{100}$, $\sigma_{010}$, $\sigma_{001}$, and $\sigma_{\bar{1}10}$, where subscripts indicate normals of the mirror planes).  In particular, the $x$-skew-stripy phase and the $y$-skew-stripy phase are related by the $\sigma_{\bar{1}10}$ symmetry, but the $z$-skew-stripy is distinct and has a different energy.  These {\it three} skew-stripy phases are local minima in the energy landscape, and the global minimum would be selected as the ground state at zero-temperature.  The energy splitting, 
\begin{align}
\Delta=(E_{z\text{-stripy}}-E_{x/y\text{-stripy}}),
\end{align}
 between these local minima as a function of $K/J$ in the skew-stripy regime is plotted in Fig. \ref{fig_diff}, with negative energies indicating a lower energy for the $z$-stripy phase.  We see that for $K/J\lesssim 8.2(\alpha\lesssim 0.8)$, the $z$-stripy phase is selected, while at $K/J=2 (\alpha=0.5)$, i.e. the exactly solvable point, the exact \textit{SU(2)} symmetry is restored and the two phases have equal energies (in fact, the quantum energy correction is identically zero at this point, since the ground state is exactly a ferromagnet in this limit).  As we shall see in the next section, there is a phase transition from the stripy-phase to a spin-liquid state at $K/J\approx 7.7(\alpha \approx 0.79)$, hence we may conclude that the $z$-skew-stripy phase is selected via quantum-order-by-disorder (QOD) in the skew-stripy regime.  We also performed the analogous analysis in the N\'{e}el regime ($J>0, K<1$): when $K>0$, QOD selects the $z$-N\'{e}el phase (spins are aligned parallel or anti-parallel to the $z$ direction), while for $K<0$, QOD selects the $x$-($y$-)N\'{e}el phase.

We conclude this section by noting that the energy splitting between $x$-/$y$- and $z$-skew-stripy phases is quite small ($\Delta\sim 10^{-6} J$), hence may be sensitive to higher-order corrections to the spin-wave spectra. More sophisticated numerical calculations based on series expansions or exact diagonalization in the future may be able to verify our present conclusion. On the other hand, we have demonstrated that the breaking of the spurious classical \textit{SU(2)} symmetry, and specifically, the lifting of degeneracy between the $x$-/$y$- and $z$-skew-phases can be achieved by considering only the lowest-order quantum corrections via spin-wave theory.

\section{Slave particle mean field theory for the Heisenberg-Kitaev Model}
\label{sec_sf}

Away from the $J=0$ limit, the Hamiltonian in Eq. \ref{eq_HK_ham} is no longer exactly solvable. In terms of the Majorana fermions the Heisenberg term is a short range four fermion perturbation. At the exactly solvable point, we find this interaction to be irrelevant at the the tree level (shown in Appendix \ref{appen_scaling}). The interactions, therefore, do not immediately destabilize the spin-liquid and a finite strength is required for causing a phase transition. This opens up a parameter regime over which the spin liquid is stable.

We study this system in the vicinity of spin liquid using a slave fermion mean field theory. As in the honeycomb case where similar calculations were done by some of the present authors,\cite{PhysRevB.86.224417} here we find it easier to work in the rotated basis (Fig. \ref{fig_rot}). 

To begin, we write the rotated spin operators as products of fermionic spinons given by\cite{PhysRevB.65.165113,PhysRevLett.58.2790} 
\begin{align}
\tilde{S}_{j}^\mu = \frac{1}{2} f_{j\alpha}^\dag [\sigma^\mu]_{\alpha \beta}f_{j\beta} .
\end{align}
 Along with the single occupancy constraint 
 \begin{align}
f^\dagger_{i\uparrow}f_{i\uparrow}+f^\dagger_{i\downarrow}f_{i\downarrow}=1,
\end{align}
this is a faithful representation of our spin Hilbert space.

Of particular interest is the portion of the phase diagram in which the Heisenberg interactions are antiferromagnetic and the Kitaev couplings are ferromagnetic. Once in the rotated basis, the rotated couplings,\cite{PhysRevB.86.224417}
\begin{align}
J'=-J;~~~K'=K-2J
\end{align}
both become ferromagnetic for $K>2J$. Due to the purely ferromagnetic couplings, we only consider spinon-hopping and pairing fields in the triplet channels which are respectively given by $\vec{E}_{ij}$ and $\vec{D}_{ij}$.\cite{PhysRevB.80.064410} Accordingly, we introduce auxiliary fields defined as
\begin{align}
E^{a}_{ij}=\langle f^\dagger_{i\alpha}\left[\tau^a\right]_{\alpha\beta}f_{j\beta}\rangle^*;~~~~D^a_{ij}=\langle f_{i\alpha}\left[i\tau^2\tau^a\right]_{\alpha\beta}f_{i\beta}\rangle^*;
\label{eq_tripans}
\end{align}
($a=x,y,z$) on each bond. In addition to these we include a magnetic decoupling 
\begin{align}
m_j = \frac{1}{2} \langle f_{j\alpha}^\dag [\sigma^z]_{\alpha \beta}f_{j\beta}\rangle
\end{align}
 which allows us to capture the magnetic ordering, which we take to be in the z direction. Choice of this direction is motivated by our semiclassical results in the previous section.

The mean-field spinon Hamiltonian takes the form
\begin{align}
H^{MF} &= \sum_{\langle ij \rangle} -\vec{f}_i{}^\dag U_{ij} \vec{f}_j \\\nonumber
 &- \frac{1}{8}J'm ( f_{i,\alpha}^\dag [\sigma^z]_{\alpha \beta} f_{i,\beta}+ f_{j,\alpha}^\dag [\sigma^z]_{\alpha \beta} f_{j,\beta} ),\\
\vec{f}_i^\dag &= \begin{bmatrix}
f_{i, \uparrow}{}^\dag & f_{i,\downarrow} & f_{i,\uparrow}^\dag & -f_{i,\downarrow}
\end{bmatrix},\nonumber
\end{align}
and the matrix $U_{ij}$ is given by
\begin{align}
U_{ij} = \sum_r& a_r \sigma^r \left(E_{ij}^r (\tau^0+\tau^3) + E_{ij}^{r\ast}(\tau^0-\tau^3)\right) \nonumber \\
+& a_r \sigma^r \left( -D_{ij}^r\tau^- + D_{ij}^{r\ast}\tau^+ \right),
\end{align}
where $\sigma^r$ are Pauli matrices acting on the spin degrees of freedom, $\tau^r$ are Pauli matrices acting on the gauge degrees of freedom and $a_r = \frac{1}{16}(J' + (1-\delta_r)K')$ with $\delta_r$ = 1 if $ij$ is an $r$ bond, and $\delta_r$ = 0 otherwise.\cite{PhysRevB.86.224417}

To continue, we choose an ansatz with translational invariance of the mean field operators, consistent with the form of the exact solution. Upon doing so, we can perform a Fourier transformation of our full mean field Hamiltonian, resulting in
\begin{align}
H^{MF} &= \sum_k \vec{\alpha}_k^\dag H_k \vec{\alpha}_k, \nonumber \\
\vec{\alpha}_k{}^\dag &= \begin{bmatrix}
\vec{f}_{k}{}_1^\dag & \vec{f}_{k}{}_2^\dag & \vec{f}_{k}{}_3^\dag & \vec{f}_{k}{}_4^\dag
\end{bmatrix}, \nonumber \\
\vec{f}_{k}{}_\beta^\dag &= \begin{bmatrix}
f_{k\beta \uparrow}^\dag & f_{-k\beta \downarrow} & f_{k\beta \downarrow}^\dag & -f_{-k\beta \uparrow}
\end{bmatrix}, \nonumber \\
H_k &= \begin{bmatrix}
m P & U_1 & 0 & A_k \\
U_1^\dag & m P & B_k & 0\\
0 & B_k^\dag & m P & U_3 \\
A_k^\dag & 0 & U_3^\dag & 0 \\
\end{bmatrix},
\end{align}
with $A_k = U_4 e^{-i{\bf k}\cdot {\bf a}_1} + U_5 e^{-i{\bf k}\cdot {\bf a}_2}$, $B_k = U_2 + U_6 e^{-i{\bf k}\cdot {\bf a}_3}$ and $P = \frac{3}{8}J'\times diag(-1,-1,1,1)$. The matrices $U_\alpha$ refer to the matrices $U_{ij}$ defined on the inequivalent links.

This leaves us with a mean field theory consisting of 37 complex parameters. We can simplify this considerably by enforcing the symmetries of the lattice. Before doing so, we note that the mean field ground state solution need not, in general, obey all of the symmetries of the lattice since spinons transform only under projective symmetries.\cite{PhysRevB.65.165113} However, here, we consider the case where the symmetries are manifest in the spinon Hamiltonian.  We also note that these symmetry operations may only relate the parameters up to gauge transformations. The presence of the inversion symmetry allows us to relate the magnitudes of the parameters on the two $z$-bonds to one another, and the three $C_2$ symmetries allow us to relate the magnitudes of the parameters on the $x$ and $y$-bonds. A self consistent mean field analysis on this model finds that stable non-dimerized solutions exist which obeys the above conditions.

\begin{figure*}
  \centering
  \subfigure[ The spinon band structure at the exactly solvable point $K/J=\infty$ ($\alpha=1$).]{
    \setlength\fboxsep{0pt}
    \setlength\fboxrule{0.0pt}
    \fbox{\includegraphics[scale=.1,clip=true,trim=0 0 0 0]{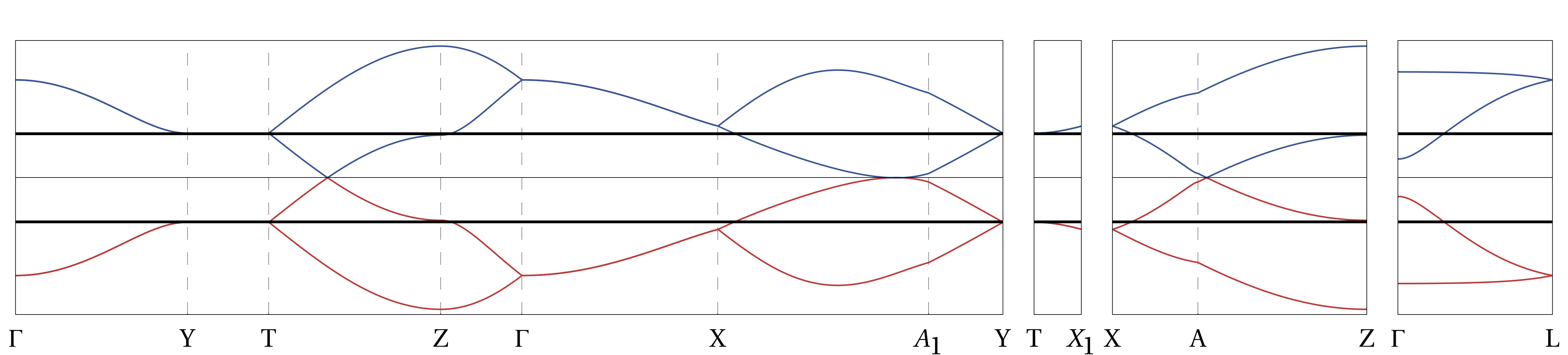}}
    \label{fig_spinon_pda}
  }
  \subfigure[ The spinon band structure at the point $K/J=8$ ($\alpha=0.8$). At this point, magnetic order has not yet stabilized.]{
    \setlength\fboxsep{0pt}
    \setlength\fboxrule{0.0pt}
    \fbox{\includegraphics[scale=.1,clip=true,trim=0 0 0 0]{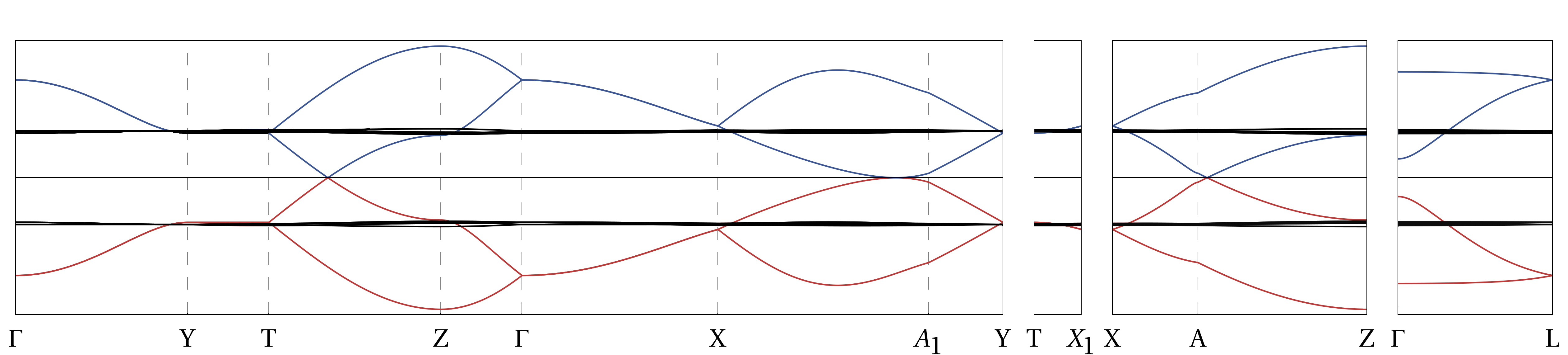}}
    \label{fig_spinon_pdb}
  }
  \subfigure[ The spinon band structure at the point $K/J=4.9$ ($\alpha=0.71$). At this point, magnetic order is present.]{
    \setlength\fboxsep{0pt}
    \setlength\fboxrule{0.0pt}
    \fbox{\includegraphics[scale=.1,clip=true,trim=0 0 0 0]{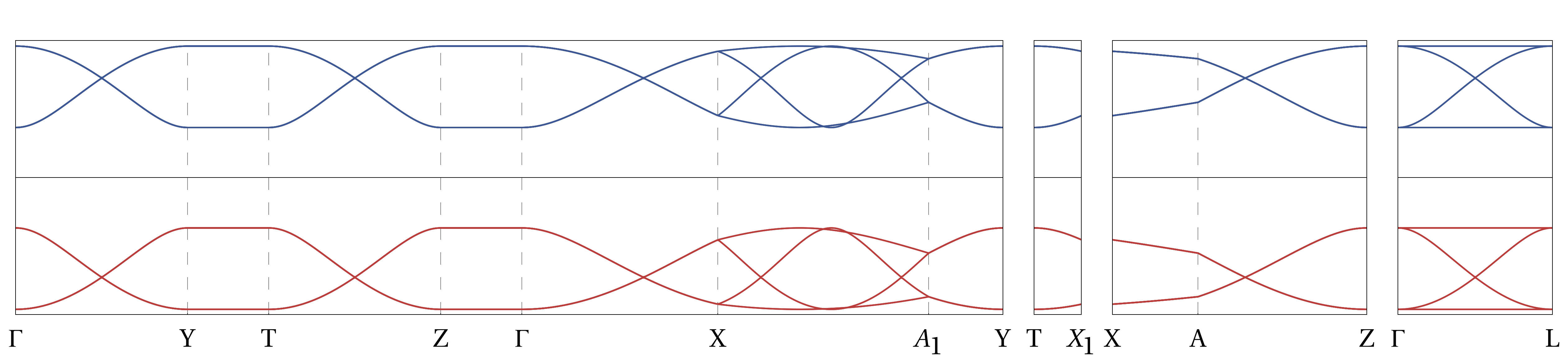}}
    \label{fig_spinon_pdc}
  }
  \caption{}
  \label{fig_spinon_pd}
\end{figure*}

Further insight into the spin liquid can be obtained from considering the relation to the exact solution in the $J$=0 limit. Choosing the form of the mean field parameters to be
\begin{align}
D_{ij}^x, E_{ij}^z &\in {\rm{Imaginary}},\nonumber\\
 D_{ij}^y &\in {\rm{Real}},
\label{eq_signstruct}
\end{align}
with the remaining components set to zero, we can diagonalize our mean field Hamiltonian in terms of Majorana fermions. We use the basis described by You {\it et al.}\cite{PhysRevB.86.085145} defining our four Majorana modes as 
\begin{align}
&c_i=\frac{1}{\sqrt{2}}(f_{i\uparrow} + f_{i\uparrow}^\dag );~~~~b^x_i=\frac{1}{i\sqrt{2}}(f_{i\downarrow} - f_{i\downarrow}^\dag )\nonumber\\
&b_i^y=\frac{-1}{\sqrt{2}}(f_{i\downarrow} + f_{i\downarrow}^\dag );~~~~b_i^z=\frac{1}{i\sqrt{2}}(f_{i\uparrow} - f_{i\uparrow}^\dag ).
\label{eq_majorana}
\end{align}
Performing a self-consistent mean field theory in terms of these parameters, we find that our minimum energy non dimerized solution is consistent with the symmetries discussed above. This can be related to the exact solution as discussed by Schaffer {\it et al.}\cite{PhysRevB.86.224417} In particular, we find that the parameters have the values $E^z = -0.11603i$, $D^x = -iD^y = 0.38397i$ on $z$-bonds, $D^x = -0.12443i$, $E^z = -iD^y = 0.37557i$ on $x$-bonds, and $D^y = 0.12443$, $E^z = D^x = 0.37557i$ on $x$-bonds. This anisotropy between the mean field parameters on the $z$-bonds compared to the $x$ and $y$-bonds is due to the absence of a symmetry relating these bond types as discussed previously.

Examining the spinon dispersion in this limit, we find that we have four dispersing fermion modes which reproduce the features of the exact solution, in addition to 12 flat bands (see Fig. \ref{fig_spinon_pda}). The flat bands are not fully degenerate, due to the differences between the mean field parameters on the $z$ and $x,y$-bonds. The spin liquid is gapless, with a Fermi surface at the zone boundary similar to the exact solution described above.

As we move away from the $J=0$ limit, we keep the structure of the mean field parameters as described by Eq. \ref{eq_signstruct}, while allowing the values of these to evolve. We also reintroduce the magnetization order parameter $m$, to capture the competing order to the spin liquid. As we begin to perturb away from the point $J=0$, we find that the spinon bands which were previously flat gain a dispersion, with an energy which scales with $J$, as shown in Fig. \ref{fig_spinon_pdb}. These bands remain fully gapped, and although they do not contribute to the low energy theory they do cause further neighbour spin correlations to become non-zero.\cite{PhysRevB.84.155121} The location of the Fermi surface changes slightly as we perturb away from this limit, but it maintains its structure; it remains a single line node on the zone boundary as in Fig. \ref{fig_fermi}.

\begin{figure}
\setlength\fboxsep{0pt}
\setlength\fboxrule{0.pt}
\fbox{\includegraphics[scale=.09,clip=true,trim=0 0 -50 0]{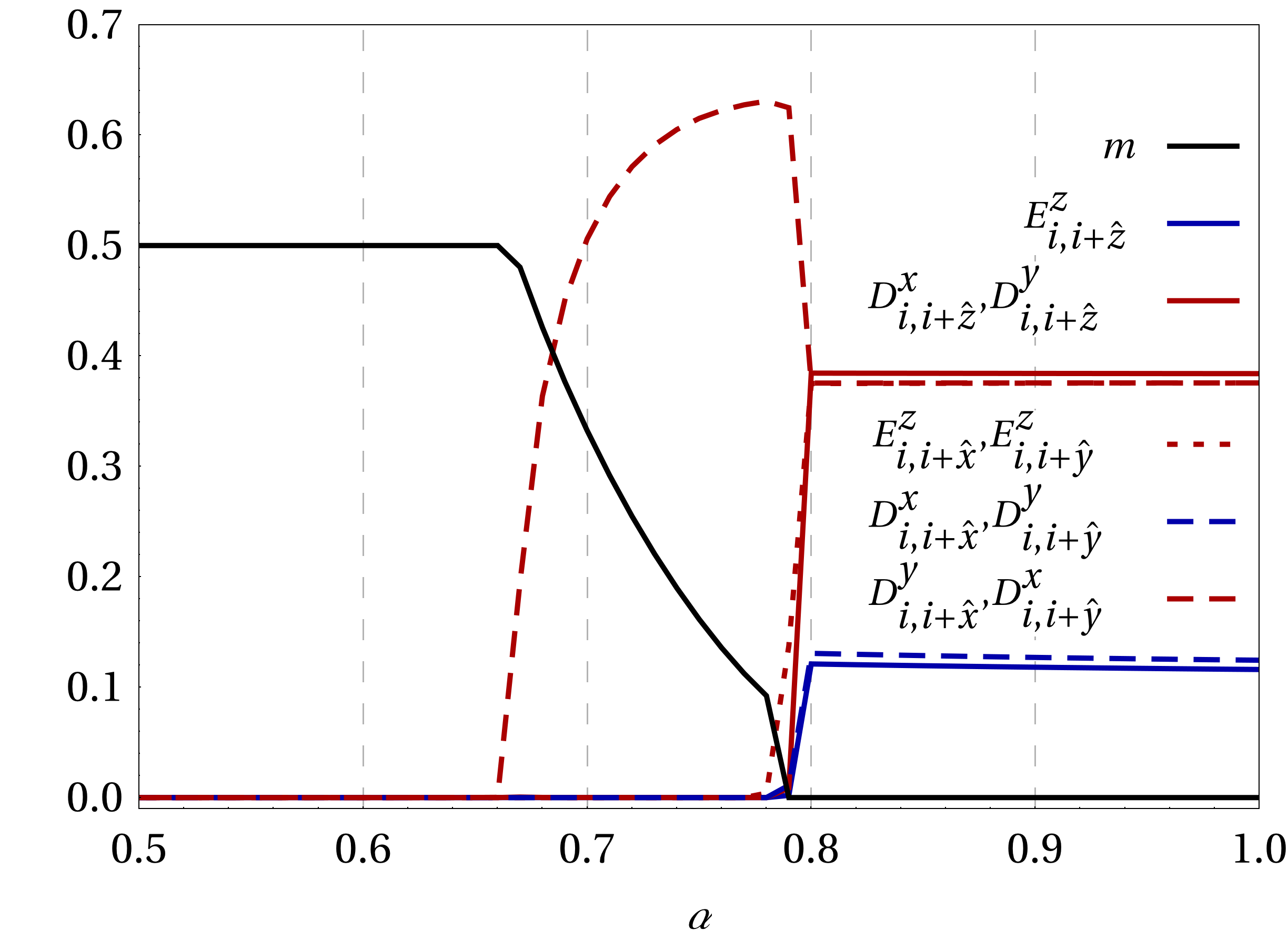}}
\label{fig_MFparams}
\caption{The magnitude of the mean field order parameters, plotted as a function of $\alpha$, where $J=(1-\alpha)$ and $K=2\alpha$}
\end{figure}

As we increase $J$, the mean field theory finds a first order phase transition into a phase with non-zero net magnetization (in the rotated basis). The transition occurs at approximately $K/J \approx 7.7~(\alpha\approx 0.79)$ (see Fig. \ref{fig_MFparams}). This transition significantly alters the spinon band structure, resulting in the formation of a gap as well as a significant change of the general structure, as shown in Fig. \ref{fig_spinon_pdc}. As we increase the value of $J$, all of the hopping and pairing amplitudes are driven to zero, and the model becomes fully described by the stripy magnetic ordering.

\section{Response to magnetic field and finite temperature}
\label{sec_magt}

In this section, we briefly discuss the finite field and finite temperature effects in the skew-stripy and the spin liquid phases.

\subsection{$h\neq 0, T\neq 0$ effect on the skew-stripy phase}

When an external magnetic field is applied perpendicular to the $x$, $y$, or $z$ directions (see Fig. \ref{fig_lattice} and Appendix \ref{appen_hyperhoneycomb} for definition of these directions), the magnetic response of the skew-stripy phase can be computed analytically at the classical level.  We highlight that the saturation field is \textit{only} dependent on the Heisenberg exchange and \textit{not} on the Kitaev coupling.  

To see this, we minimize the classical energy functional at zero temperature.  First, we write our spin-configuration as a sum of a variational component, $\psi$, and a ferromagnetic component along our applied field, $\psi_{\rm FM}$ 
\begin{align}
\Psi=\sqrt{(1-\bar m^2)}\cdot\psi+\bar m\cdot\psi_{\text{FM}}.
\end{align}
Similar to the Luttinger-Tisza method, the variational component $\psi$ is subject to the constraint that every spin in $\Psi$ must have the same length.  We have also introduced the variational parameter $\bar m~(|\bar m| \leq 1)$ as the (dimensionless) magnetization along our applied field.  As a result, the energy functional can written as
\begin{equation}
  E([\Psi],\bar m)=e[\psi]\cdot(1-\bar m^2)+e[\psi_{\text{FM}}]\cdot \bar m^2-h\cdot \bar m
\end{equation}
where $\vec{h}=h \hat{h}$ is the external field and $e[\psi](e[\psi_{\rm FM}])$ are the energies of the variational (ferromagnetic) components of the wave function.

In the skew-stripy regime ($K>J>0$) and for $\hat{h}$ perpendicular to $x$, $y$, or $z$, minimizing the energy with respect to $\psi$ subject to the aforementioned constraint gives $\psi=\psi_{\text{skew-stripy}}$ in the $x$, $y$, or $z$ direction respectively.  Substituting the energies per spin of the skew-stripy and ferromagnetic states into the energy functional, we obtain
\begin{equation}
  E(\bar m)=\frac{-J-K}{8}(1-\bar m^2)+\frac{3J-K}{8}\bar m^2-h \bar m.
\end{equation}
Minimizing in respect to $\bar m$ yields the relation $\bar m=h/J$, i.e. magnetization saturates at $h_{\text{sat}}=J$, which is independent of the Kitaev coupling $K$.  We contrast this with the N\'{e}el regime ($J>0, K/J<1$), where the above analysis will yield $h_{\text{sat}}=(3J-K)/2$.

Though the above result is purely classical, it is nonetheless quite interesting since it suggests that the magnetic field response is controlled by only the Heisenberg parameter, $J$. The insensitivity of the saturation field to the Kitaev coupling $K$ in the skew-stripy regime may provide a useful tool to probe the value of the ratio $K/J$ for the actual material.

Turning to the finite temperature response, we immediately note that the mean-field Curie-Weiss temperature is given by:\cite{PhysRevB.84.100406}
\begin{align}
\Theta_{\rm CW}=\frac{1}{4}(K-3J).
\end{align}
So, for $K>3J$, $\Theta_{CW}>0$ as in the honeycomb case. Further, while both the energy scales, $J$ and $K$, enters into the expression for $\Theta_{\rm CW}$, only the former, as seen before, enters into the saturation value of the magnetic field. 

The low energy magnetic specific heat, at the quadratic level, receives major contributions from the quadratically dispersing spin-wave mode near the $Y$-point (Fig. \ref{fig_spinwaves_stripy}). This leads to a specific heat that is proportional to $T^{3/2}$. This power-law is expected to be cut-off at a temperature scale that corresponds to the gap of the mode (when higher order magnon-magnon interactions are taken into account). 
\subsection{$h\neq 0, T\neq 0$ effect on the spin liquid}

In zero magnetic field, the spin-spin correlations at the pure Kitaev point are strictly nearest neighbour.\cite{PhysRevLett.98.247201} For finite $J$, the spin-spin correlations are exponentially decaying.\cite{PhysRevB.84.155121} On putting in a magnetic field, we expect this to change to a power-law similar to the honeycomb case.\cite{PhysRevLett.106.067203}

The low temperature specific heat in the spin liquid regime is controlled by the gapless fermions. Since the spinon band-gap vanishes on a one-dimensional manifold, the low temperature magnetic specific heat scales as $\sim T^2$ (shown in Appendix \ref{appen_scaling}). 
\section{Discussion and Outlook}
\label{sec_discuss}

In summary, motivated by recent experiments by Takagi {\it et. al} on $\beta$-Li$_2$IrO$_3$,\cite{2013_takagi} we have studied the possibility of realizing a Heisenberg-Kitaev spin model on the hyperhoneycomb lattice (Fig. \ref{fig_lattice}). We argue that the spin physics of this material in the strong coupling limit, where Ir$^{4+}$ ions carrying localized $J_{\rm}=1/2$ moments surrounded by edge sharing oxygen octahedra with Ir-O-Ir bond angle being 90$^\circ$, may be essentially captured by a Heisenberg-Kitaev model in three dimensions. Using a combination of semiclassical analysis, exact solution and slave-fermion mean field theory, we study the phase diagram of this model that allows interesting magnetically ordered phases as well as an extended window of a three dimensional gapless Z$_2$ spin liquid phase. In among the magnetically ordered phases, in addition to the usual N\'{e}el and the ferromagnet, we find two other collinear phase--the skew-stripy and the skew-zig-zag. Focusing on the antiferromagnetic Heisenberg-Ferromagnetic Kitaev regime ($J,K>0$ in Eq. \ref{eq_HK_ham}), we find that the quantum fluctuations select the $z$-skew stripy phase as the energy minimum through quantum-order-by disorder. The spin liquid, on the other hand, has gapless Fermi-circles (Fermi-surface with co-dimensions, $d_c$=2). This occur at the Brillouin zone boundary and has interesting implications at low temperature. Our slave-fermion mean-field theory predicts a first order transition between the spin liquid and the magnetically ordered skew-stripy phase. 

In regards to actual experiments on the material, it would be interesting to see if any of the above phases are relevant to describe the physics of actual material $\beta$-Li$_2$IrO$_3$. We predict the general form of the low temperature specific heat and also the magnetic field dependence for the susceptibility in both the skew-stripy and the spin liquid regimes. Interestingly, in the classical limit, the magnetic field required to saturate the system only depends on the magnitude of the Heisenberg coupling ($J$), while the Curie-Weiss temperature contains both Heisenberg ($J$) and Kitaev ($K$) couplings. This may indicate that the temperature response and the magnetic field response, particularly the magnetization saturation, energy scale may be quite different. These results can be compared with respect to future experiments. The spin-wave spectra can similarly be compared to future neutron scattering studies on this compound. Overall, the possibility of realizing another family of Mott insulators where the Heisenberg-Kitaev model is relevant would be exciting with the possibility of realizing a three dimensional quantum spin liquid phase that this model allows.

\acknowledgements
We thank K. Hwang and H. Takagi for discussions. YBK wishes to acknowledge the hospitality of MPIPKS, Dresden. This research was supported by the NSERC, CIFAR, and Centre for Quantum Materials at the University of Toronto.  After submission of this manuscript to the pre-print arXiv, an independent and related work\cite{kimchi2013} with some overlapping results appeared soon after.
\appendix

\begin{figure}
  \centering
  \setlength\fboxsep{0pt}
  \setlength\fboxrule{0.0pt}
  \fbox{\includegraphics[scale=.8,clip=true,trim=0 0 0 0]{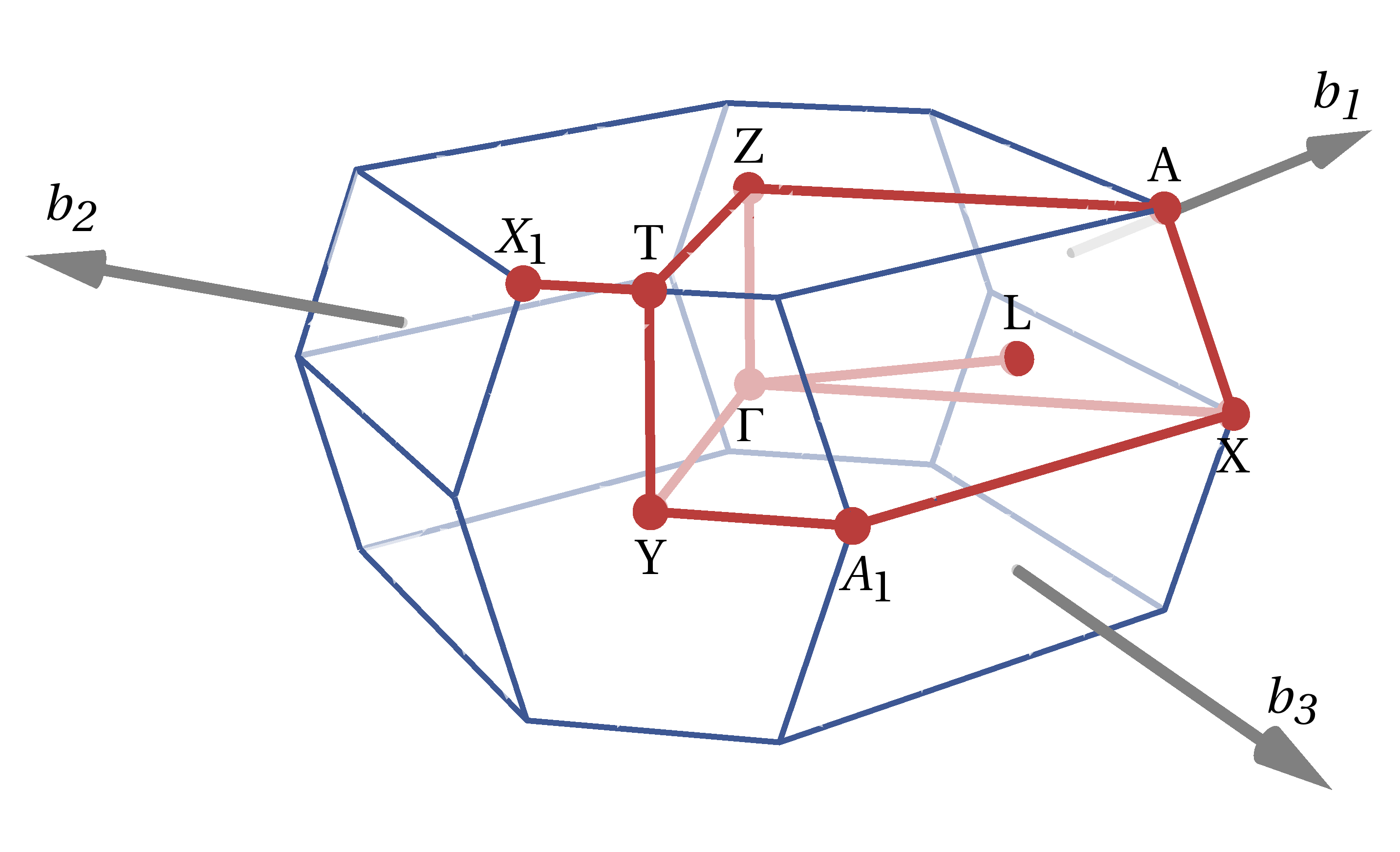}}
\caption{The Brillouin zone. The high symmetry paths are: $\Gamma\rightarrow Y\rightarrow T\rightarrow Z\rightarrow\Gamma\rightarrow X\rightarrow A_1\rightarrow Y$; $T\rightarrow X_1$; $X\rightarrow A\rightarrow Z$ and $\Gamma\rightarrow L$. The following are the position of the high symmetry points: $\Gamma=(0,0,0)$, $Y=\left(0,0,-\frac{\pi }{2}\right)$, $T=\left(-\frac{\pi }{6},-\frac{\pi }{6},-\frac{\pi }{2}\right)$, $Z=\left(-\frac{\pi }{6},-\frac{\pi }{6},0\right)$, $X=\left(\frac{29 \pi }{72},-\frac{29 \pi }{72},0\right)$, $A_1=\left(\frac{11 \pi }{72},-\frac{11 \pi }{72},-\frac{\pi }{2}\right)$, $X_1=\left(-\frac{19 \pi }{72},-\frac{5 \pi }{72},-\frac{\pi }{2}\right)$, $A=\left(\frac{13 \pi}{72},-\frac{37 \pi}{72},0\right)$ and $L=\left(\frac{\pi }{6},-\frac{\pi }{3},-\frac{\pi }{4}\right)$.}
\label{fig_bz}
\end{figure}

\section{\label{appen_hyperhoneycomb}The structure of the ideal hyperhoneycomb lattice}

Here we elaborate on the lattice structure of the ideal hyperhoneycomb.  The ideal structure has $90\,^{\circ}$ Ir-O-Ir bonds, $120\,^{\circ}$ Ir-Ir-Ir bonds, and perfect oxygen octahedra around each Ir$^{4+}$ ion. All nearest-neighbour Ir-Ir bonds have the same length. The lattice can be described by a face-centerd orthorhombic lattice with a four site basis. The primitive face-centered orthorhombic lattice vectors are given by
\begin{align}
  \mathbf{a_1}=(2,4,0),~~ 
  \mathbf{a_2}=(3,3,2),~~ 
  \mathbf{a_3}=(-1,1,2).
\end{align}
This choice of lattice vectors, shown in Fig. \ref{fig_lattice},  ensures that both Ir and O ions have positions possessing integer coordinates.  For instance, the four Ir ions now have the positions
\begin{align}
  \text{Ir}_1=(0,0,0),~ 
  \text{Ir}_2=(1,1,0),~
  \text{Ir}_3=(1,2,1),~
  \text{Ir}_4=(2,3,1)
\end{align}
and the 6 oxygens around each Ir are located at $\pm\mathbf{\hat{x}}$,$\pm\mathbf{\hat{y}}$, and $\pm\mathbf{\hat{z}}$ relative to the Ir position.  We also note that the oxygen ions form a face-centered orthorhombic lattice by themselves in the ideal hyperhoneycomb.  

One can also describe the lattice structure with the enlarged orthorhombic unit cell as illustrated in Fig. \ref{fig_lattice}.  In this case the lattice vectors are given by $\mathbf{a}=(6,6,0)$, $|a|=6\sqrt{2}$, $\mathbf{b}=(-2,2,0)$, $|b|=2\sqrt{2}$, and $\mathbf{c}=(0,0,4)$, $|c|=4$ (in the same units as those used in the above lattice vectors).
\subsection{The first Brillouin zone}

The reciprocal lattice vectors are given by:
\begin{align}
{\bf b}_1&=\left(\frac{\pi }{3},-\frac{2 \pi }{3},\frac{\pi }{2}\right),\nonumber\\
{\bf b}_2&=\left(-\frac{2 \pi }{3},\frac{\pi }{3},-\frac{\pi }{2}\right),\nonumber\\
{\bf b}_3&=\left(\frac{2 \pi }{3},-\frac{\pi }{3},-\frac{\pi }{2}\right).
\end{align}
The first Brillouin zone as well as the high symmetry directions and points are shown in Fig. \ref{fig_bz}.

\section{Choice of the link variables $u^\alpha_{ij}$ in the zero flux sector and the zero-flux hopping Hamiltonian}
\label{appen_zeroflux}

\begin{figure}
\centering
  \centering
  \setlength\fboxsep{0pt}
  \setlength\fboxrule{0.0pt}
  \fbox{\includegraphics[scale=.1,clip=true,trim=50 250 0 0]{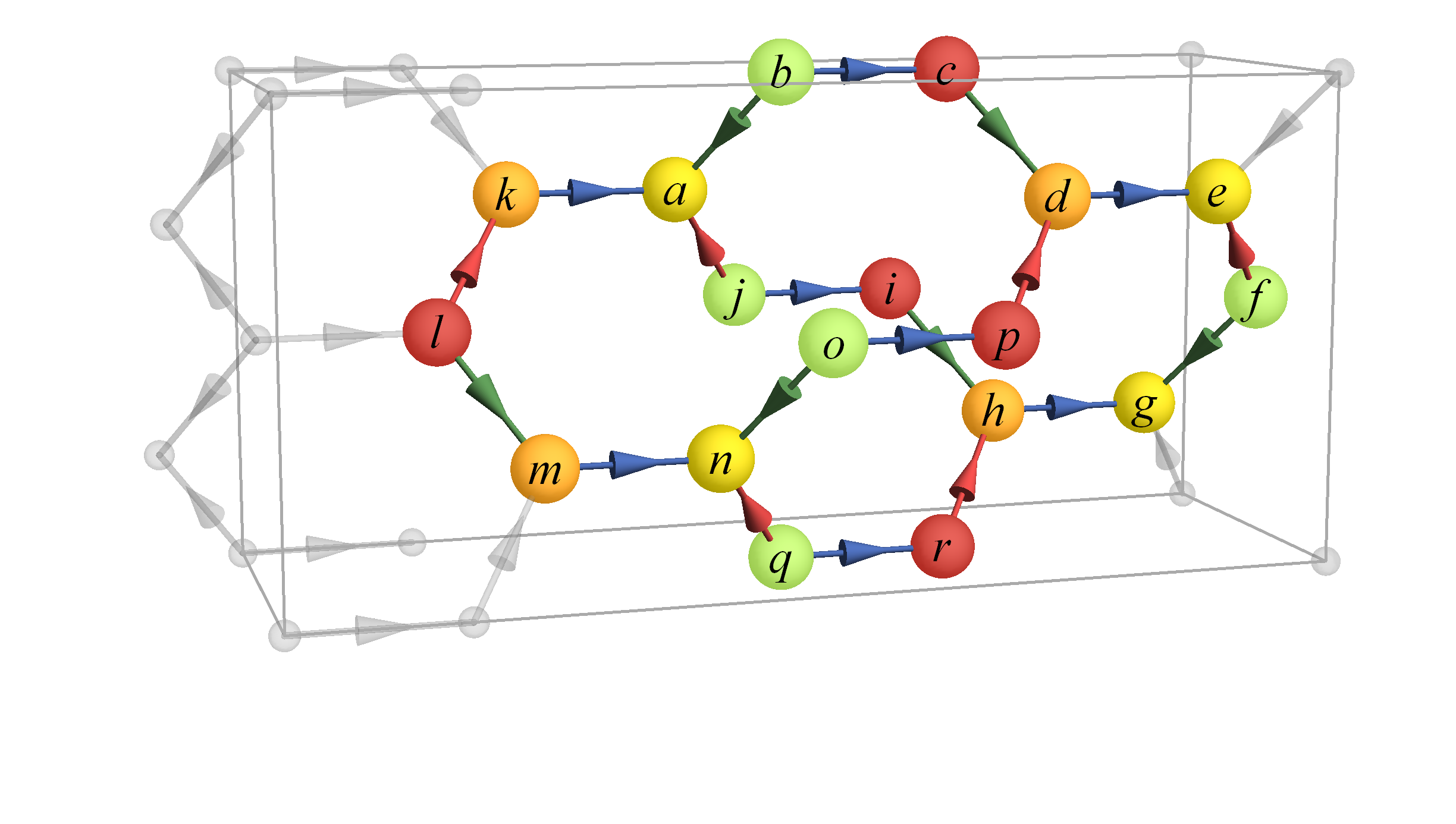}}
\caption{The 4 loops as shown are: (1) b-c-d-e-f-g-h-i-j-a; (2) m-n-o-p-d-c-b-a-k-l; (3) m-n-q-r-h-i-j-a-k-l; (4) q-r-h-g-f-e-d-p-o-n.  Sublattices 1, 2, 3, and 4 are colored green, red, orange, and yellow respectively to aid visualization of the four-site unit cell.}
\label{fig_loops}
\end{figure}

Defining the loop variables in terms of the spins, according to the discussion following Eq. \ref{eq_spinflux}, for the four kinds of 10-site loops (shown in Fig. \ref{fig_loops}) we have
\begin{align}
\mathcal{W}_{P_1}&=2^{10}S^{x}_{b}S^{x}_{c}S^{x}_{d}S^{y}_{e}S^{z}_{f}S^{x}_{g}S^{x}_{h}S^{x}_{i}S^{y}_{j}S^{z}_{a}\\
\mathcal{W}_{P_2}&=2^{10}S^{x}_{m}S^{x}_{n}S^{x}_{o}S^{y}_{p}S^{z}_{d}S^{x}_{c}S^{x}_{b}S^{x}_{a}S^{y}_{k}S^{z}_{l}\\
\mathcal{W}_{P_3}&=2^{10}S^{x}_{m}S^{y}_{n}S^{y}_{q}S^{y}_{r}S^{z}_{h}S^{x}_{i}S^{y}_{j}S^{y}_{a}S^{y}_{k}S^{z}_{l}\\
\mathcal{W}_{P_4}&=2^{10}S^{y}_{q}S^{y}_{r}S^{y}_{h}S^{x}_{g}S^{z}_{f}S^{y}_{e}S^{y}_{d}S^{y}_{p}S^{x}_{o}S^{z}_{n}
\end{align}
Using
\begin{align}
\sigma^x&=-i\sigma^y\sigma^z=-\imath b^yb^z\\
\sigma^y&=-i\sigma^z\sigma^x=-\imath b^zb^x\\
\sigma^z&=-i\sigma^x\sigma^y=-\imath b^xb^y
\end{align}
where $S^\alpha=\sigma^\alpha/2$ and $\sigma^\alpha~~(\alpha=1,2,3)$ are the Pauli matrices. For the four loops we then get
\begin{align}
\mathcal{W}_{P_1}&=\uij{z}{bc}\uij{y}{cd}\uij{z}{de}\uij{x}{fe}\uij{y}{fg}\uij{z}{hg}\uij{y}{ih}\uij{z}{ji}\uij{x}{ja}\uij{y}{ba}\\
\mathcal{W}_{P_2}&=\uij{z}{mn}\uij{y}{on}\uij{z}{op}\uij{x}{pd}\uij{y}{cd}\uij{z}{bc}\uij{y}{ba}\uij{z}{ka}\uij{x}{lk}\uij{y}{lm}\\
\mathcal{W}_{P_3}&=u^{z}_{mn}u^{x}_{qn}u^{z}_{qr}u^{x}_{rh}u^{y}_{ih}u^{z}_{ji}u^{x}_{ja}u^{z}_{ka}u^{x}_{lk}u^{y}_{lm}\\
\mathcal{W}_{P_4}&=u^{z}_{qr}u^{x}_{rh}u^{z}_{hg}u^{y}_{fg}u^{x}_{fe}u^{z}_{de}u^{x}_{pd}\uij{z}{op}u^{y}_{on}u^{x}_{qn}
\end{align}
On a loop, therefore, if we choose a gauge where the above link variables are $+1$ then we are in the zero flux sector. This is shown in Fig. \ref{fig_loops} where $u^\alpha_{ij}=+1$ when going from $i$ to $j$ we traverse along the arrow. Further, this configuration of $u^\alpha_{ij}$ has the same unit cell as the lattice and so one can use the 4-site unit cell for diagonalization.

Now in this zero flux sector, the hopping Hamiltonian is given by
Eq. \ref{eq_zerohop} where, as stated in the main text, $ij$ are given by the direction of the arrows in Fig. \ref{fig_loops}.  Therefore, we can write it more explicitly as:
\begin{align}
H^{0-\rm flux}_{\rm K}=\frac{\imath}{2}\sum_{\bf R}\left[c_{1,\bf R}\left(c_{2,\bf R}+c_{4,\bf R-a_1}+c_{4,\bf R-a_2}\right)\right.\nonumber\\
\left.+c_{3,\bf R}\left(c_{4,\bf R}-c_{2,\bf R}-c_{2,\bf R+a_3}\right)\right]
\end{align}


\section{The tree level scaling for short-range four fermion interactions for Fermi surface with co-dimensions, $d_c=2$ in three spatial dimensions}
\label{appen_scaling}

In the Kitaev model, we have both dispersing Majorana fermions, $c_j$ as well as ones which have a flat band, $b^\alpha_j~~(\alpha=x,y,z)$. 

For dispersing fermions in $d$ spatial dimensions, where the Fermi surface has a co-dimension of $d_c(<d)$, the free action is\cite{PhysRevLett.102.046406}
\begin{align}
\mathcal{S}_{0,c}=\int d\omega\int d^{d-d_c}{\bf l}\int d^{d_c}{\bf k}{c}_{\bf k,l}\left(i\omega-v_{\bf l}\cdot{\bf k}\right)c_{\bf k,l}
\end{align}
where the ``directions" denoted by ${\bf l}$ lie on the Fermi surface and hence do not scale while ${\bf k}$ denotes the direction away from the Fermi sruface\cite{RevModPhys.66.129,PhysRevLett.102.046406} For the flat band fermions, the schematic form of the action is given by
\begin{align}
\mathcal{S}_{0,b}=\int d\omega\int d^{d-d_c}{\bf l}\int d^{d_c}{\bf k}b_{\bf k,l}\left(i\omega-\epsilon_0\right)b_{\bf k,l}
\end{align}
(where we have suppressed the superscript $\alpha$ which is not important for the present calculation)

Using the scaling
\begin{align}
\omega'=\lambda\omega\nonumber\\
{\bf l}'=\lambda^0{\bf l}\nonumber\\
{\bf k}'=\lambda{\bf k}\nonumber\\
\end{align}
(where $\lambda>1$ is the scaling parameter) we get:
\begin{align}
c'_{\bf k',l'}&=\lambda^{-\frac{d_c+2}{2}}c_{\bf k,l}\\
b{^\alpha}_{\bf k,l}'&=\lambda^{-\frac{d_c+1}{2}}b^\alpha_{\bf k,l}
\end{align}
The Heisenberg interactions are typically given by:
\begin{align}
\mathcal{S}_4=g\int \left[\prod_{i=1}^3 d\omega_i d^{d-d_c}{\bf l}_id^{d_c}{\bf k}_i \right]b_{\bf k_1,l_1,\omega_1}b_{\bf k_2,l_2,\omega_2}c_{\bf k_3,l_3,\omega_3}c_{\bf k_4,l_4,\omega_4}
\end{align}
where $A_4=-(A_1+A_2+A_3)~~(A={\bf k},{\bf l},\omega)$. Using the scaling at the pure Kitaev point, we find that
\begin{align}
[g]=-d_c
\end{align}
Hence the four fermion interaction of the Heisenberg type is irrelevant at the Kitaev point ($d_c=2$). We just note that this is more irrelevant than the four fermion vertex which is of $cccc$ type. This latter vertex has a scaling dimension of $1-d_c$.
\subsection{The scaling of the low temperature specific heat}
The low temperature specific heat receives contribution from the $c$ fermions. It is given by:
\begin{align}
C&\sim \frac{\partial}{\partial T}\int d^{d-d_c}{\bf l}\int d^{d_c}{\bf k}\frac{|{\bf k}|}{e^{|{\bf k}|/T}+1}\sim T^{d_c}
\end{align}

\bibliography{biblio}
\end{document}